\documentclass[twocolumn]{aastex62}

\usepackage[normalem]{ulem}
\usepackage{color}

\defcitealias{Chevalier85}{CC85}

\begin{document}

\title{Production of Cool Gas in Thermally-Driven Outflows}

\correspondingauthor{Evan Schneider}
\email{es26@astro.princeton.edu}

\author[0000-0001-9735-7484]{Evan E. Schneider\footnotetext{}}
\altaffiliation{Hubble Fellow}
\affiliation{Department of Astrophysical Sciences, Princeton University, 4 Ivy Lane, Princeton, NJ 08544, USA}

\author[0000-0002-4271-0364]{Brant E. Robertson}
\affiliation{Department of Astronomy and Astrophysics, University of California, Santa Cruz, 1156 High Street, Santa Cruz, CA 95064, USA}

\author{Todd A. Thompson}
\affiliation{Department of Astronomy and Center for Cosmology and AstroParticle Physics, Ohio State University, 140 W 18th Ave, Columbus, OH, USA}

\begin{abstract}
Galactic outflows commonly contain multiphase gas, and its physical origin requires explanation. Using the CGOLS (Cholla Galactic OutfLow Simulations) suite of high-resolution isolated galaxy models, we demonstrate the viability of rapid radiative cooling as a source of fast-moving ($v \sim 1000$ km/s), cool ($10^4$ K) gas observed in absorption line studies of outflows around some star-forming galaxies. By varying the mass-loading and geometry of the simulated winds, we identify a region of parameter space that leads to cool gas in outflows. In particular, when using an analytically-motivated central feedback model, we find that cooling flows can be produced with reasonable mass-loading rates ($\dot{M}_{wind} / \dot{M}_{SFR} \sim 0.5$), provided the star formation rate surface density is high. When a more realistic clustered feedback model is applied, destruction of high density clouds near the disk and interactions between different outflow regions indicate that lower mass-loading rates of the hot gas within the feedback region may still produce multiphase outflows. These results suggest an origin for fast-moving cool gas in outflows that does not rely on directly accelerating cool gas from the interstellar medium. These cooling flows may additionally provide an explanation for the multiphase gas ubiquitously observed in the halos of star-forming galaxies at low redshift.
\end{abstract}


\section{Introduction}\label{sec:intro}

Both observations and cosmological simulations indicate that feedback from star formation plays an important role in galaxy evolution. In particular, winds driven by supernovae are invoked in order to explain the stellar mass function of low-mass galaxies \citep{Dekel86, Mashchenko08, Peeples11}, as well as the mass-metallicity relation \citep{Tremonti04,Erb08,Finlator08}, and the metal enrichment of the circumgalactic medium (CGM) \citep{Madau01, Scannapieco02, Oppenheimer06}. By transporting gas out of galaxies, outflows limit the fuel for future star formation. If the outflowing gas is significantly enriched by supernovae and moving with high velocity, it can pollute metal-poor gas in the CGM and beyond.

Observations support this picture via detections of outflowing gas around star-forming galaxies at all redshifts. Blue-shifted absorption in the spectra of both individual galaxies and stacks indicates that gas is being driven out of galaxies at a range of velocities and temperatures \citep{Weiner09, Steidel10, Erb12, Kornei12, Martin12, Bordoloi14, Rubin14, Heckman15}. Recent studies using the Cosmic Origins Spectrograph (COS) on the Hubble Space Telescope have probed the CGM of local galaxies at distances out to the virial radius \citep[see review by][]{Tumlinson17}. These surveys have found neutral hydrogen with high covering fractions in the halos of both star-forming and passive galaxies, indicating the ubiquity of cool gas in galactic halos at low redshift \citep{Prochaska11,Tumlinson13,Werk14,Werk16,Borthakur16}. Intriguingly, the absorption lines from cool gas are systematically stronger and have larger widths and velocity offsets around starburst systems as compared to average star-forming galaxies \citep{Heckman17b}.

While the prevalence of galactic winds has led to many investigations of their properties, a detailed understanding of the physics that drives outflows and governs their evolution remains elusive. From an observational perspective, many studies provide only a single data point for a given galaxy, and a host of assumptions must be made about the outflow location, geometry, and ionization state in order to estimate the total amount of gas in a particular phase \citep{Murray07,Chisholm17}. On the theoretical side, \textit{ab initio} simulations of supernovae-driven winds are computationally prohibitive, requiring resolution of order $\sim1\,\mathrm{pc}$ to capture the generation of winds within the galaxy and volumes spanning tens or hundreds of kpc to track their evolution.

The challenges associated with observing and simulating galactic winds have resulted in a number of outstanding questions about their nature. In particular, ``down-the-barrel" absorption-line studies have revealed the presence of fast-moving ($500 - 1000\,\mathrm{km}\,\mathrm{s}^{-1}$), cool ($10^4\,\mathrm{K}$) gas in many systems \citep{Veilleux05,Heckman17}. Simulations and analytic studies have demonstrated that such gas is difficult to produce via ram pressure acceleration of interstellar medium (ISM) material. Rather than accelerating, high density clouds tend to be rapidly shredded and incorporated into the hot phase of the wind \citep{Scannapieco15, Bruggen16, Schneider17, Zhang17}, though caveats remain regarding the role of magnetic fields in stabilizing the clouds \citep{McCourt15, Banda-Barragan18} or additional sources of driving such as radiation pressure on dust grains \citep[e.g.,][]{Murray05, Thompson15, Zhang18}, or cosmic rays \citep[e.g.,][]{Breitschwerdt91, Socrates08, Girichidis16}. Given the uncertain origin of cool gas in both outflows and the CGM, recent theoretical work has explored the possibility of producing cool gas from the hot phase of the winds via thermal instability, and high resolution, local box simulations have suggested the viability of such an origin \citep{Scannapieco17}.

In 1985, \citeauthor{Chevalier85} \citepalias[hereafter][]{Chevalier85} published an analytic description of a starburst-driven wind. The \citetalias{Chevalier85} model describes an expanding hot wind driven by a constant mass and energy injection within a spherical region. By solving the spherical equations of hydrodynamics, solutions for various parameters of interest can be found as a function of radius, including the density, velocity, pressure, and mach number of the wind. The \citetalias{Chevalier85} model has been used to successfully explain numerous observations of starburst galaxies, particularly the hot plasma at the center of the nearby nuclear starburst galaxy, M82 \citep{Strickland09}.

Despite its utility as a model for the hot phase of a galactic outflow, the \citetalias{Chevalier85} model does not address potentially large energy losses due to radiative cooling, limiting its applicability once the hot wind has reached a radius at which these physical processes become dynamically important. The \citetalias{Chevalier85} model alone cannot explain the presence of multiphase gas in an outflow, which is often observed in starburst systems \citep[e.g.,][]{Heckman00,Veilleux05}. Following the work of \cite{Wang95} and \cite{Silich04}, \cite{Thompson16} recently explored an extension to the \citetalias{Chevalier85} model that includes the effects of radiative energy losses from the hot phase of the outflow as it expands \citep[see also][]{Bustard16}. Briefly, an outflow is expected to undergo rapid cooling once it reaches a radius where the cooling time roughly equals the advection time. For systems with particularly compact star-forming regions, \cite{Thompson16} propose that these rapidly-cooling, mass-loaded outflows could be the origin of the cool ($\sim10^{4}\,\mathrm{K}$) gas observed in absorption around many star-forming systems as both outflowing winds and galactic-wind-driven CGM shells.

Rapid cooling of the hot phase of galactic winds would prove an intriguing origin for the multiphase gas observed in many starburst systems, but modeling this formation mechanism numerically requires high resolution over a relatively large volume. The resolution of the simulations and the feedback models employed will set the scale of the density inhomogeneities that may lead to thermal instability within the flow, and the cool clouds that form may have sizes much smaller than the system as a whole. In addition, high resolution is required to capture hydrodynamic instabilities that affect the distribution of density, velocity, and temperature of gas in a multiphase outflow once it has formed. In an effort to make progress on this front, we use the hydrodynamics code Cholla \citep{Schneider15} to run a series of galactic wind simulations with unprecedented resolution and a variety of feedback models. The simulations are collectively called the ``Cholla Galactic OutfLow Simulations'' (CGOLS) suite \citep{Schneider18}. CGOLS is designed to increase systematically in complexity from one simulation to the next, allowing clean comparisons to the analytic models of outflows at each step in order to pin down the physical origin of different outflow features.

In this paper, we seek to test the idea of cooling in galactic outflows. In doing this, we numerically validate the 1D analytic model of \cite{Thompson16} and test an asymmetric ``clustered" 3D outflow geometry that cannot be well-modeled analytically. In the following section, we describe the numerical setup for the CGOLS simulations, including the parameter space of the analytic model to be explored, the initial conditions for our isolated galaxy setup, and the feedback prescriptions employed. We then move on in Section~\ref{sec:results} to a description of the results, primarily focusing on the large-scale properties of the outflow as a function of radius (Section \ref{sec:radial_structure}). We briefly touch on the velocity structure and multiphase nature of the outflow seen in the clustered feedback model in Sections \ref{sec:velocity_structure} and \ref{sec:multiphase}, but reserve more detailed discussion of the multiphase gas for a subsequent paper (Schneider et al. \textit{in prep}). Section~\ref{sec:discussion} contains a discussion of our results in the context of other simulations and observations of outflows. We conclude and discuss directions for future work in Section~\ref{sec:conclusions}.

\section{Methods} \label{sec:methods}

In the following subsections, we describe the galactic wind model as explored by the CGOLS suite. First, we outline the analytic model of radiatively-cooling winds. We then give a brief description of the setup of our numerical simulations, which are more thoroughly outlined in an accompanying paper \citep{Schneider18}. Details of the hydrodynamical model as implemented in the novel GPU-based Cholla code are included at the end of this section, and we refer the interested reader to the Cholla code paper for further information \citep{Schneider15}. 

\subsection{The Analytic Model for Cooling Outflows}

Following the derivation presented in Section 2.1 of \cite{Thompson16}, we characterize the analytic model of cooling winds with three parameters: the radius of the injection region, $R$, the mass injection rate, $\dot{M}_\mathrm{wind}$, and the energy injection rate, $\dot{E}_\mathrm{wind}$. The mass injection rate into the wind is described as a function of the star formation rate according to the formula $\dot{M}_\mathrm{wind} = \beta \dot{M}_\mathrm{SFR}$, where $\beta$ is the ``mass-loading factor"\footnote{In this paper, we refer to ``mass-loading" of the wind to indicate mass that has been incorporated into the hot phase at some point. We refer to ``entrainment'' to describe gas from the disk that has been lofted into the outflow, but has not undergone a phase transition.}. Similarly, the energy injection rate can be related to the star formation rate by making assumptions about supernova rate and energy. Specifically, if each supernova is assumed to release $10^{51}\,\mathrm{erg}$ of energy and there is 1 supernova per $100\,\mathrm{M}_\odot$ of stars formed, then $\dot{E}_\mathrm{wind} = 3\times10^{41}\,\mathrm{erg}\,\mathrm{s}^{-1} \alpha \dot{M}_\mathrm{SFR}$, where the factor $\alpha$ accounts for the fraction of the supernova energy that is thermalized in the hot plasma and not radiated away by dense gas \added{, and $\dot{M}_\mathrm{SFR}$ is measured in solar masses per year}.

By assuming a simple power-law function for the cooling rate as a function of temperature, \cite{Thompson16} derive two characteristic timescales for the hot wind: the cooling time, 
\begin{equation}
t_\mathrm{cool} \simeq 3\times10^6\,\mathrm{yr}\frac{\alpha^{2.20}}{\beta^{3.20}}\left(\frac{R_{0.3}}{r_{10}}\right)^{0.27}\frac{R_{0.3}^2}{\dot{M}_{\mathrm{SFR},10}}\Omega_{4\pi},
\end{equation}
and the advection time,
\begin{equation}
t_\mathrm{adv} \simeq 1\times10^7\,\mathrm{yr}\left(\frac{\beta}{\alpha}\right)^{1/2}r_{10},
\end{equation}
where $R_{0.3} = R/0.3\,\mathrm{kpc}$ is a characteristic radius for the injection region, $r_{10} = r/10\,\mathrm{kpc}$ is a characteristic radius for the outflow, $\dot{M}_{\mathrm{SFR},10} = \dot{M}/10\,\mathrm{M}_\odot\,\mathrm{yr}^{-1}$ is a characteristic star formation rate, and $\Omega_{4\pi} = \Omega/4\pi\,\mathrm{str}$ is the opening angle of the outflow. By setting $t_\mathrm{cool} = t_\mathrm{adv}$, \cite{Thompson16} derive a ``cooling radius" for the outflow, beyond which it becomes strongly radiative:
\begin{equation}
r_\mathrm{cool} \simeq 4\,\mathrm{kpc} \frac{\alpha^{2.13}}{\beta^{2.92}}\mu^{2.13}R_{0.3}^{1.79}\left(\frac{\Omega_{4\pi}}{\dot{M}_{\mathrm{SFR},10}}\right)^{0.789}.
\label{eqn:r_cool}
\end{equation}
Here we have dropped the factor $\xi$ from the \citet{Thompson16} derivation by assuming the metallicity of the gas is solar ($\xi = 1$), and included the dependence on the mean molecular weight, $\mu$, which was assumed to be unity throughout the \citet{Thompson16} derivation. The cooling radius derived in Equation \ref{eqn:r_cool} is an estimate based on a simplified cooling function, and derived in the $r >> R$ limit, but nevertheless gives us a range of $\alpha$ and $\beta$ parameter space to explore in simulating cooling outflows. In the simulations presented below, we vary $\alpha$ and $\beta$ in a physically-motivated way, in order to test both the accuracy with which Equation \ref{eqn:r_cool} can predict the location of the cooling radius and the way in which asymmetric feedback changes the nature of the winds.

\subsection{The Simulation Suite}

The CGOLS suite (to-date) consists of three high-resolution simulations, each with $2048\times2048\times4096$ volume elements, as well as lower-resolution simulations designed to test the effects of numerical resolution, feedback parameterization, and other physics. The box size for the simulations is $10\times10\times20\,\mathrm{kpc}^{3}$, giving a fixed resolution of $\Delta x = 4.9\,\mathrm{pc}$ throughout the \textit{entire} simulation volume. This extremely high numerical resolution allows us to study the multiphase nature of the outflows that result in an unprecedented level of detail. Each simulation starts with an isothermal gas disk in rotational dynamic and vertical hydrostatic equilibrium, embedded within an initially-adiabatic hydrostatic spherical halo. After $5\,\mathrm{Myr}$, we begin to inject mass and energy according to one of two feedback prescriptions - ``central" or ``clustered". The first simulation, model `A', is an adiabatic reference simulation, with no radiative cooling. The second, model `B', is identical to model A, but with radiative cooling. The third, model `C', is a clustered, asymmetric feedback simulation, which also includes the effects of radiative cooling.

\subsubsection{Galaxy Initial Conditions}

Our initial conditions are chosen to create a disk galaxy loosely modeled after the gas-rich nearby starburst, M82. The galaxy contains $M_\mathrm{gas} = 2.5\times10^9\,\mathrm{M}_\odot$ of gas \citep{Greco12}, distributed in an isothermal disk ($T_\mathrm{gas} = 10^4\,\mathrm{K}$), with an exponential surface density profile defined by the scale radius, $R_\mathrm{gas} = 1.6\,\mathrm{kpc}$. The pressure gradient in the radial direction is balanced by the gas rotation. We artificially truncate the gas disk at $R = 4.5\,\mathrm{kpc}$ in order to reduce potential boundary effects from the simulation box. The disk is in vertical hydrostatic equilibrium with a static gravitational potential that consists of a Miyamoto-Nagai disk \citep{Miyamoto75} plus a spherically-symmetric NFW dark matter halo \citep{Navarro96}. The disk potential is set by the stellar mass, $M_\mathrm{stars} = 10^{10}\,\mathrm{M}_\odot$ \citep{Greco12}, scale radius, $R_\mathrm{stars} = 0.8\,\mathrm{kpc}$ \citep{Mayya09}, and scale height, $z_\mathrm{stars} = 0.15\,\mathrm{kpc}$ \citep{Lim13}. The halo potential is set by a dark matter mass $M_\mathrm{vir} = 5\times10^{10}\,\mathrm{M}_\odot$, viral radius $R_\mathrm{vir} = 53\,\mathrm{kpc}$, and concentration $c = 10$, leading to a halo scale radius of $R_\mathrm{h} = R_\mathrm{vir}/c = 5.3\,\mathrm{kpc}$. 

\subsubsection{Feedback Implementation}

Models A and B are run with a very simple feedback method that we call central feedback, which is designed to mimic the assumptions made in the 1D analytic models. After the simulation has run in equilibrium for $5\,\mathrm{Myr}$, we begin to inject mass and energy into a spherical region centered on the origin, with $R = 300\,\mathrm{pc}$. We choose $300\,\mathrm{pc}$ because that radius was demonstrated via 2D hydrodynamic simulations to produce a similar wind solution to a disk injection region with radius $R = 750\,\mathrm{pc}$ \citep{Strickland09}, a good fit to the size of the starburst region in M82. The mass and energy are distributed equally at each time step over the volume of all the cells within the spherical volume $V = 4\pi R^3/3$.

For the first half of each simulation, we choose our energy and mass injection rates, $\dot{E}$ and $\dot{M}$, such that we expect to see a cooling radius within $10\,\mathrm{kpc}$. In particular, from 5 to $40\,\mathrm{Myr}$, we assume a star formation rate $\dot{M}_\mathrm{SFR} = 20\,M_{\odot}\,\mathrm{yr}^{-1}$, a mass-loading factor $\beta = 0.6$, and energy-loading factor $\alpha = 0.9$. The mass-loading rate therefore assumes that the hot wind incorporates approximately 2-3 times the expected return from supernovae and stellar winds alone, which for a continuous star formation model are expected to yield an $\dot{M} \sim 0.25\dot{M}_\mathrm{SFR}$ \citep{Leitherer99}. The energy-loading rate assumes that 90\% of the energy from supernovae is thermalized in the hot gas driving the wind, where we have assumed there is one $10^{51}\,\mathrm{erg}$ supernova per $100\,M_\odot$ of stars formed. Thus, at early times, our mass injection rate is $\dot{M} = 12\,M_\odot\,\mathrm{yr}^{-1}$, and our energy injection rate is $\dot{E} = 5.4\times10^{42}\,\mathrm{erg}\,\mathrm{s}^{-1}$. Using these parameters in Equation~\ref{eqn:r_cool} gives a cooling radius of $r_\mathrm{cool} = 2.77\,\mathrm{kpc}$, assuming a spherical outflow with opening angle $\Delta\Omega = 4\pi$ and a mean molecular weight $\mu = 0.6$, as appropriate for fully-ionized solar-metallicity gas.

After $40\,\mathrm{Myr}$, we ramp down the feedback, reducing the assumed star formation rate to $\dot{M}_\mathrm{SFR} = 5\,M_\odot\,\mathrm{yr}^{-1}$, the mass-loading to $\beta = 0.3$, and we assume all of the supernova energy is thermalized, $\alpha = 1.0$. Thus, our mass and energy injection rates at late times are given by $\dot{M} = 1.5\,M_\odot\,\mathrm{yr}^{-1}$ and $\dot{E} = 1.5\times10^{42}\,\mathrm{erg}\,\mathrm{s}^{-1}$. With these parameters, the advection time for the hot wind to reach a radius of $10\,\mathrm{kpc}$ is $t_\mathrm{adv} \sim 5.5\,\mathrm{Myr}$, which is much shorter than the cooling time, $t_\mathrm{cool} \sim 93\,\mathrm{Myr}$, so we do not expect to see a cooling radius in the hot wind during the low mass-loading state. These two states are designed to roughly mimic a picture in which some event (perhaps a merger) brings in a large quantity of gas to the galaxy, and the high surface density drives a nuclear starburst. At early times, when there is still plenty of gas in the center, the mass-loading rate is higher, whereas at later times, much of the gas has been consumed by star formation or already driven out in a wind, so the mass-loading is lower. Our particular choices for $\alpha$, $\beta$, and SFR at late times are based on the observed estimates for M82 at the present day \citep{Strickland09}. However, this is just one possible picture - our goal in this paper is not to elucidate the small-scale processes that set $\alpha$ and $\beta$, but to test the effects on the the large-scale hot wind that results from a given combination of $\alpha$ and $\beta$.

The central feedback model described above is valuable for comparison to the 1D analytic models, and may even be a reasonable approximation of very concentrated nuclear starburst systems. However, star formation is generally expected to proceed in a highly clustered manner. In M82, star clusters associated with the most recent burst of star formation are observed within the central kiloparsec of the disk \citep{Mayya08}. Therefore, we also introduce a ``clustered" feedback model to simulate a situation where star formation is slightly more spread out within the disk. Comparisons between the two models will allow us to investigate the difference between idealized, axisymmetric feedback that can be modeled analytically, and the case where the mass and energy is distributed in an asymmetric manner. In order to perform the most straightforward comparison, we make the clustered model as similar to the central model as possible. Rather than injecting mass and energy into a single sphere at the center of the galaxy, we inject mass and energy into 8 spheres, each with a radius $R_\mathrm{cluster} = 150\,\mathrm{pc}$, distributed randomly within the central 1.5 kpc of the disk, and with heights randomly distributed up to $100\,\mathrm{pc}$ above and below the disk midplane. Thus, the volume into which the mass and energy are injected remains the same as the central case, but the spatial distribution of the mass and energy input is very different. In addition to randomly assigning the location of each cluster, we move the location of all 8 clusters every $15\,\mathrm{Myr}$.

A physical rationale for the clustered feedback model exists in simulations of the superbubble mode of supernova feedback, as well as recent observations that attempt to constrain the base radius of galactic outflows. For example, \cite{Kim17} use resolved multiphase ISM simulations of clustered supernovae to estimate the shell-formation radius of a superbubble as a function of the ambient surrounding ISM density and energy input rate of the supernovae. A superbubble formed by clustered supernovae using the energy input rate assumed in our high mass-loading state would be expected to form a shell at $R\sim15\,\mathrm{pc}$, which is larger than the scale height of the ISM in the high-density region of the disk where we seed the clusters. Therefore, we assume that the bubble would break out of the disk and vent freely into the CGM, in which case all the mass and energy can reasonably be distributed evenly within the $150\,\mathrm{pc}$ sphere. The same superbubble modeling indicates that the thermalization efficiency and mass-loading factors may be similar to those we assume in this case, though \cite{Kim17} only simulated bubbles with lower ambient ISM densities and less frequent supernovae than those assumed in our simulations. From an observational perspective, \cite{Chisholm16} recently constrained the maximum radius above the disk of the mass-loading region for the outflow in NGC 6090 to be $150\,\mathrm{pc}$ - comparable to the radius of the injection region for our clustered feedback model. Thus, while our clustered model does not capture the star formation and feedback physics happening in the gas disk, it provides a reasonable model for studying the properties of the hot wind.

\subsubsection{Numerical Model}

All of the simulations in the CGOLS suite are run using Cholla \citep{Schneider15}, a highly-efficient GPU-based hydrodynamics code. Cholla includes a variety of integration methods, and the details of the methods used in this work are presented in a companion paper \citep{Schneider18}. The basics of the numerical model are included here for completeness. We employ a predictor-corrector integration method based on the description in \cite{Stone09}, along with a second-order spatial reconstruction technique with limiting applied in the characteristic variables \citep{LeVeque02} and the HLLC Riemann solver \citep{Toro94, Batten97}.

Radiative cooling is included via operator-splitting, with the thermal energy losses accounted for after the hydrodynamic update. The cooling rates are determined using a forward Euler integration, and sub-cycling of the radiative cooling time step is employed to ensure a numerically robust solution. The hydrodynamic time step is further tied to the cooling time of the gas such that no cell loses more than 10\% of its thermal energy in a given hydrodynamic time step. The cooling function is a piece-wise parabolic fit to a solar metallicity, collisional ionization equilibrium curve calculated using Cloudy \citep{Ferland13}. Cooling is only included for gas above $10^4\,\mathrm{K}$; below this cutoff a temperature floor is assumed for all the gas in the simulation. The cutoff is primarily employed to prevent collapse of the isothermal gas disk, but we note that the rarefied gas in the outflow will be subject to photo-ionization heating from the cosmic UV background, and therefore would not be expected to cool much below $10^4\,\mathrm{K}$ \citep[see][]{Schneider17}. Photo-ionization heating is not explicitly included in these simulations. All of the gas in the simulations is assumed to be ionized, with an adiabatic index $\gamma = 5/3$ and a mean molecular weight $\mu = 0.6$ used in all calculations and analysis.

\section{Results} \label{sec:results}

\begin{figure*}
\centering
\includegraphics[width=0.3\linewidth]{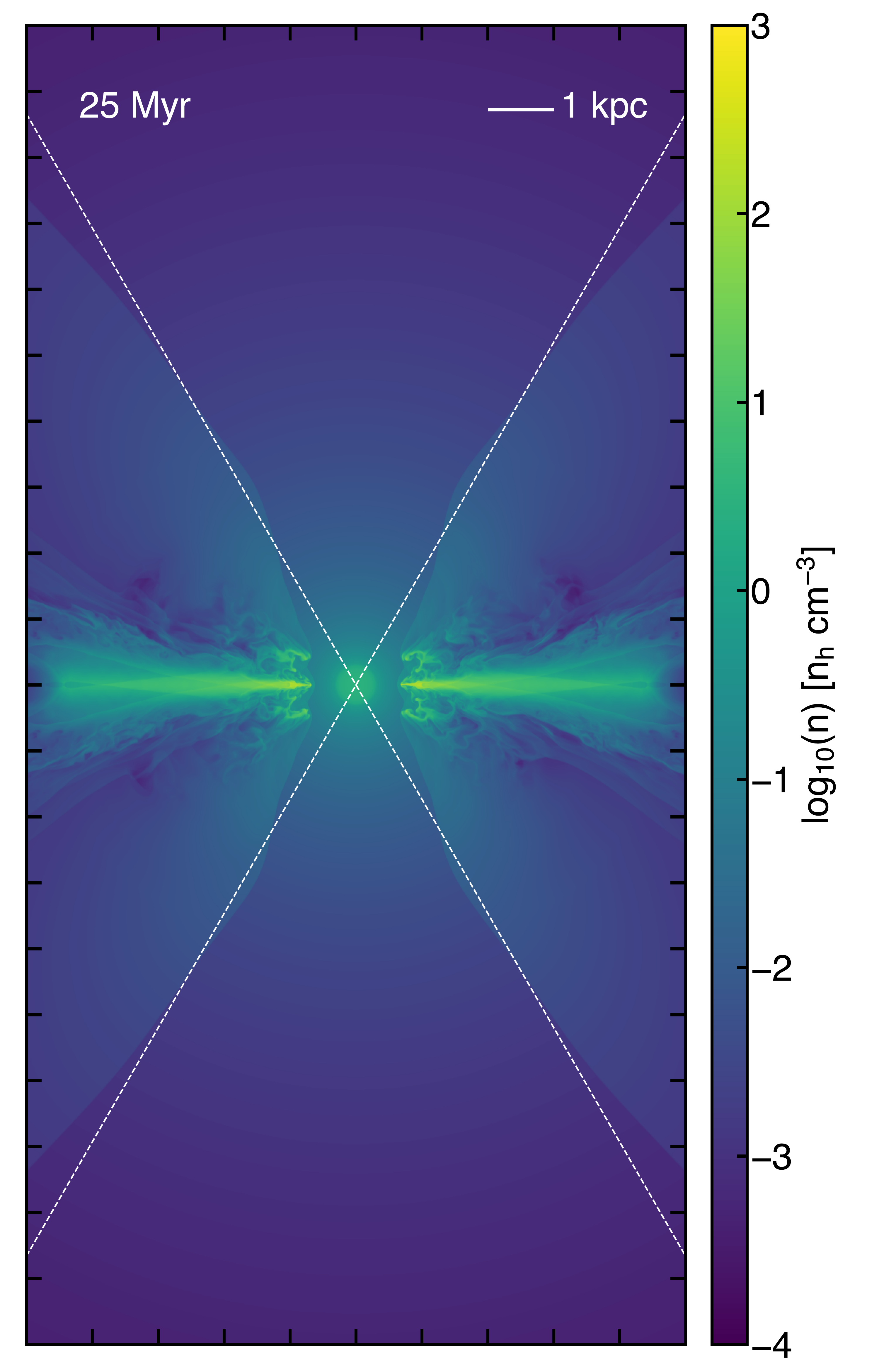}
\includegraphics[width=0.3\linewidth]{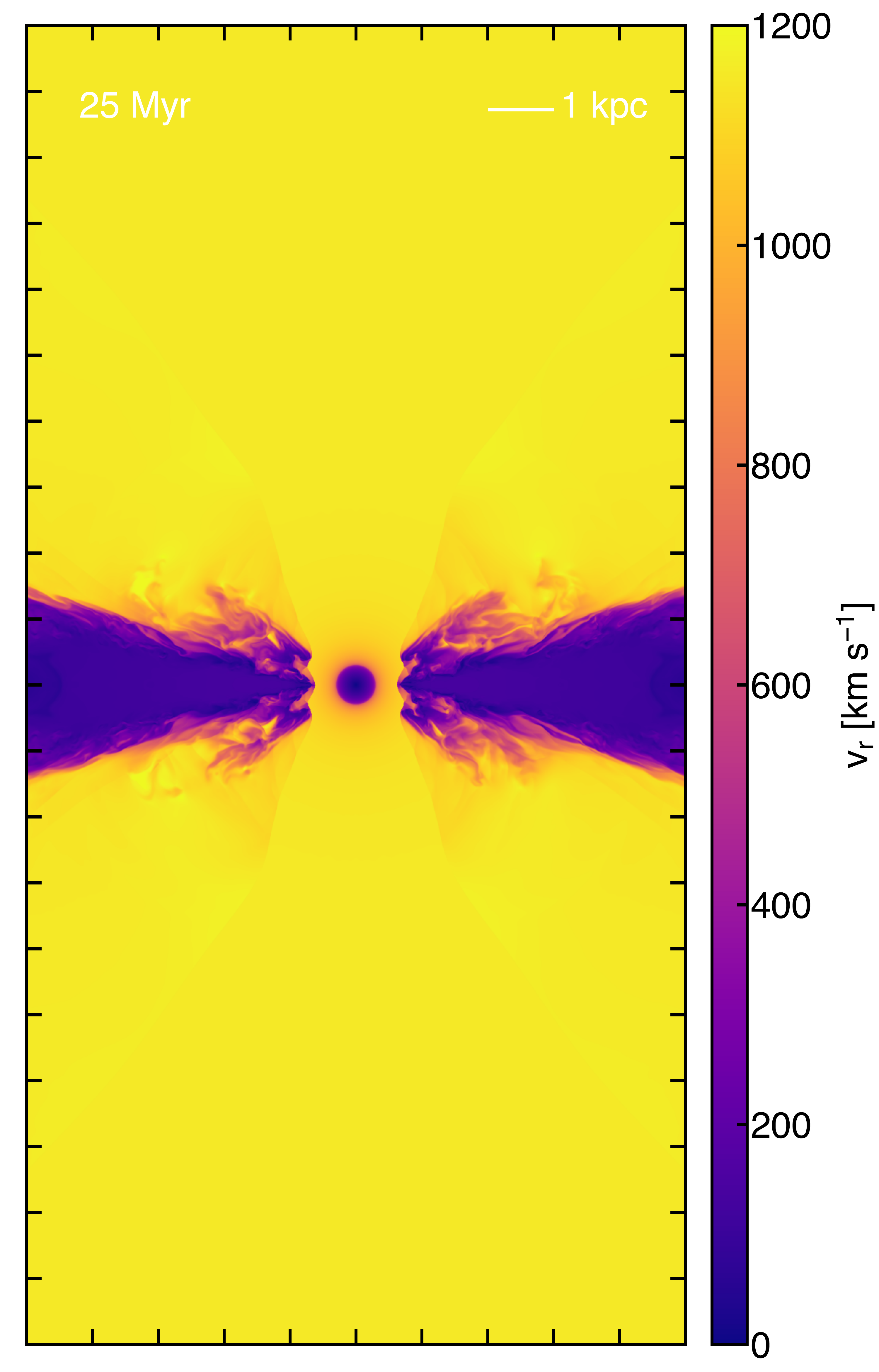}
\includegraphics[width=0.3\linewidth]{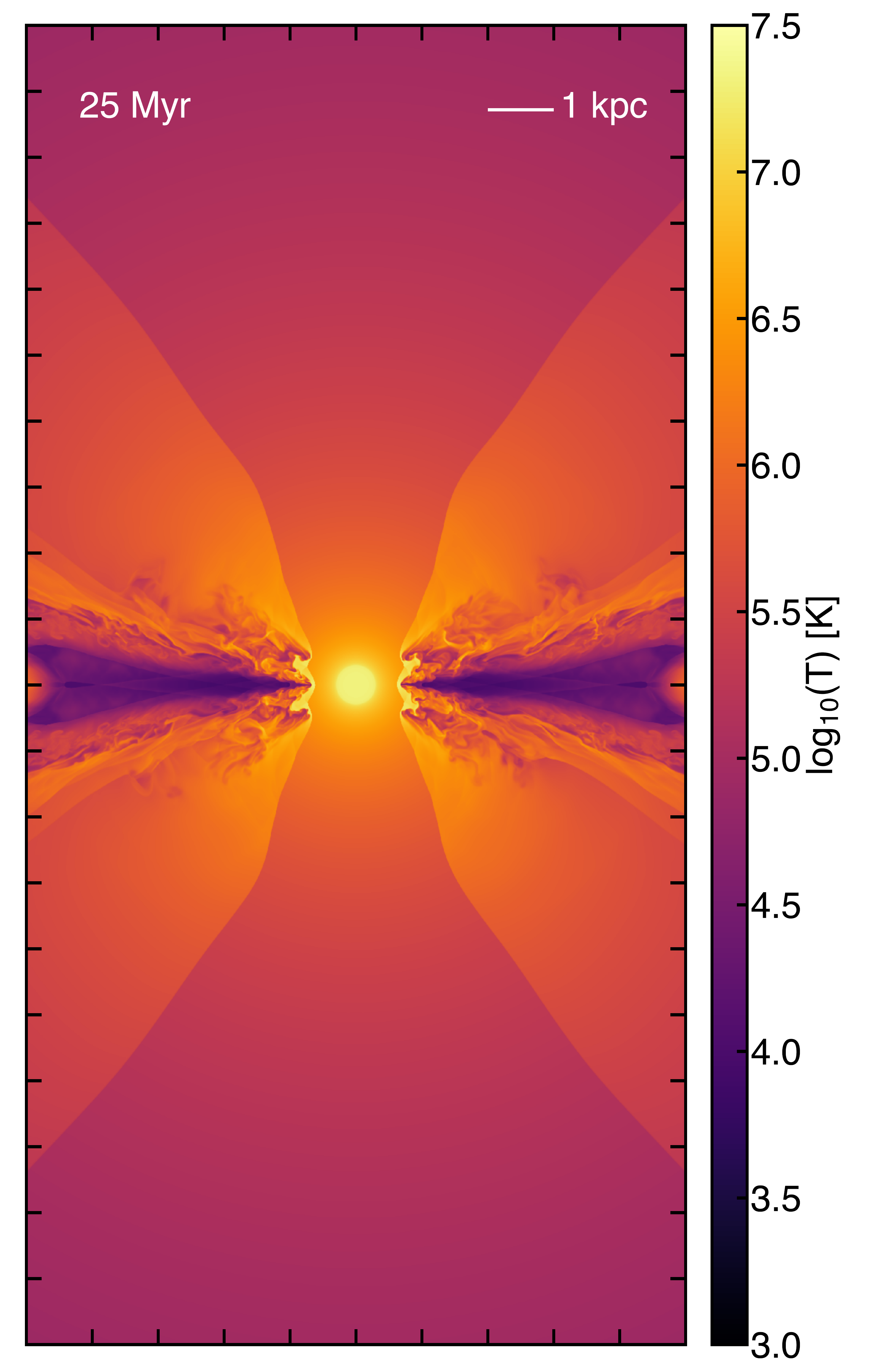}
\caption{Density\added{, radial velocity,} and temperature slices from the adiabatic, central feedback simulation (Model A) at 25 Myr. Slices are made through the $x-z$ plane, and show the isothermal disk, spherical mass and energy injection region, bi-conical free wind zone, and turbulent interface between the disk and outflow. \added{Dashed lines in the density panel show the bi-conical region with opening angle $\Delta \Omega = 60^\circ$ that will be used to calculate radial outflow properties.}}
\label{fig:adiabatic_slices_25}
\end{figure*}

\begin{figure*}
\centering
\includegraphics[width=0.3\linewidth]{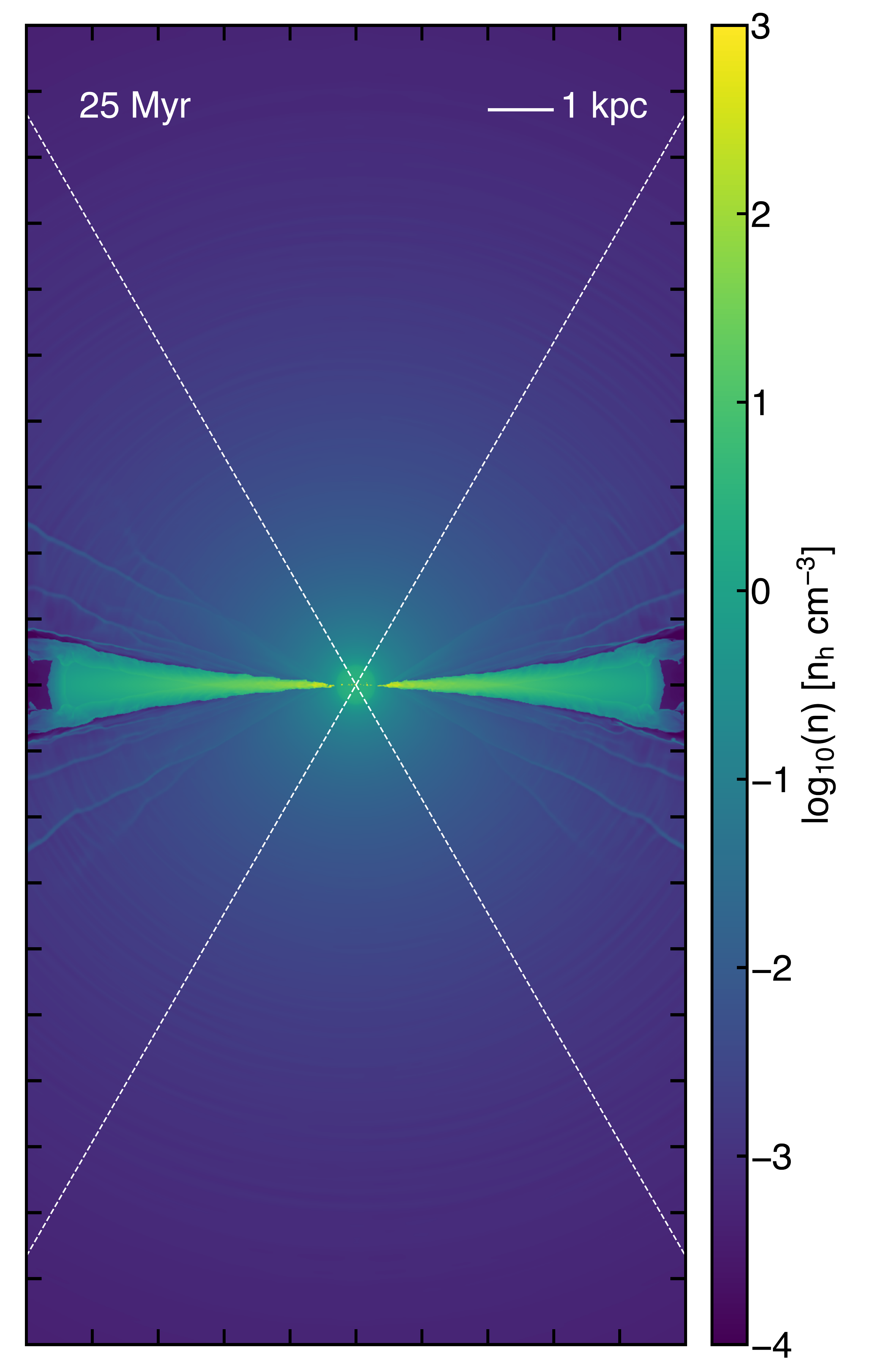}
\includegraphics[width=0.3\linewidth]{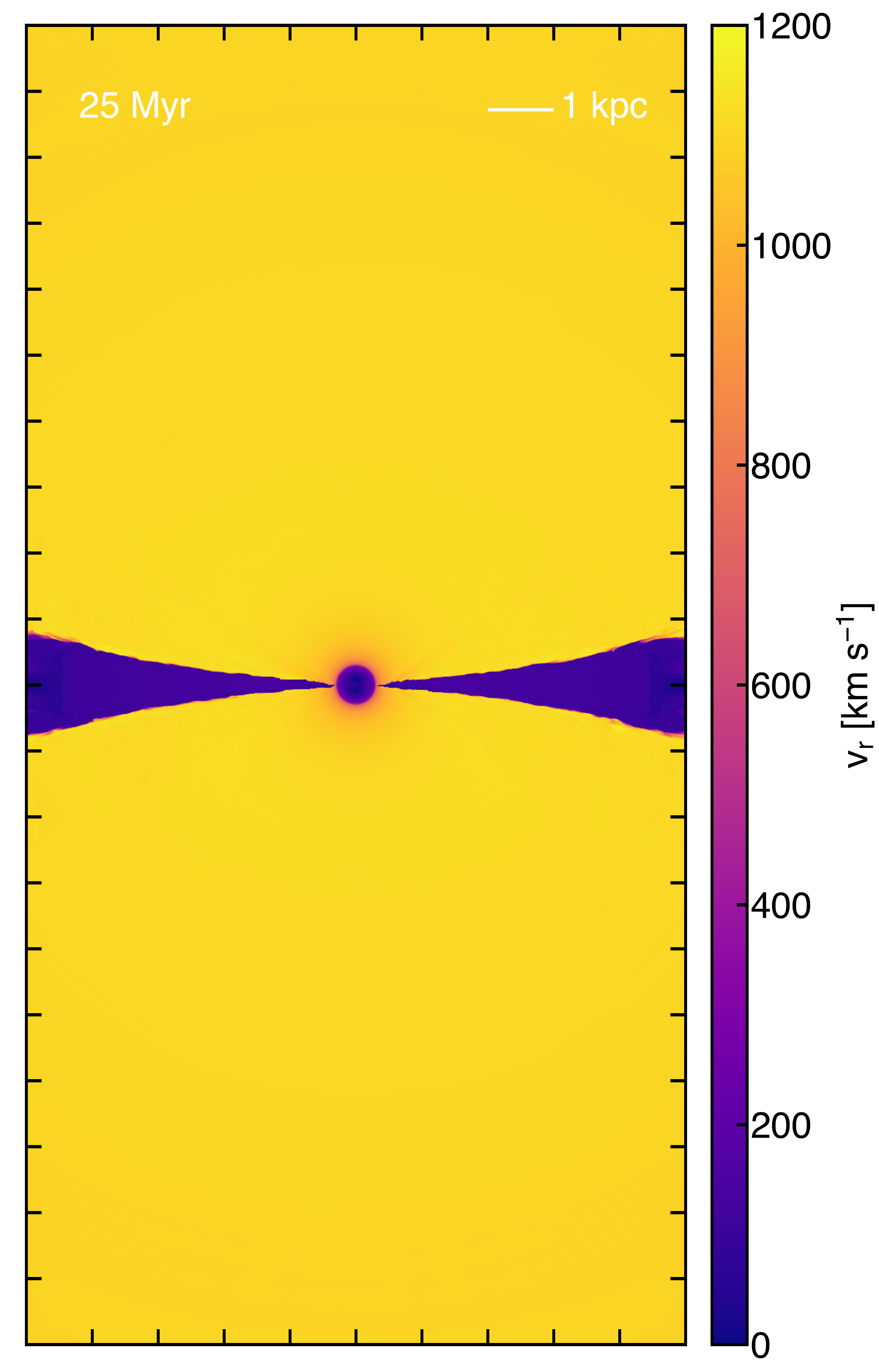}
\includegraphics[width=0.3\linewidth]{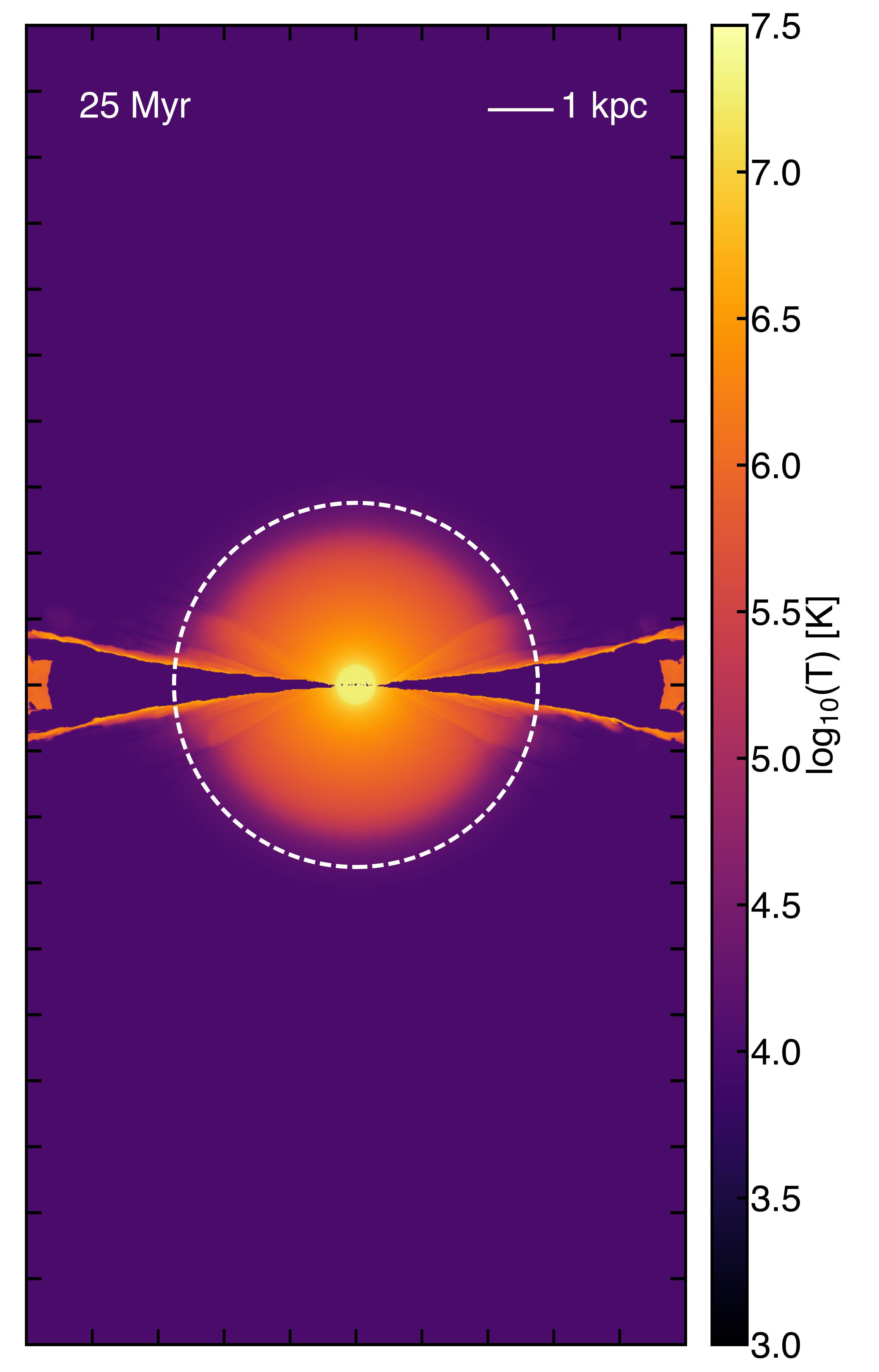}
\caption{Density\added{, radial velocity,} and temperature slices from the cooling, central feedback simulation (Model B) at $25\,\mathrm{Myr}$, during the high mass-loading outflow state. The high densities in the outflow lead to large radiative energy losses between 2 and $3\,\mathrm{kpc}$, visible as a cooling radius in the temperature slice. Beyond this radius all gas in the outflow remains at the simulation temperature floor of $10^4\,\mathrm{K}$. The analytically-predicted cooling radius (Equation~\ref{eqn:r_cool}) is plotted as a white dashed circle.}
\label{fig:cooling_slices_25}
\end{figure*}

We begin our results by presenting a phenomenological description of the three high-resolution simulations: the central feedback model without cooling (Model~A), the central feedback model with cooling (Model~B), and the clustered feedback model with cooling (Model~C). As explained in Section \ref{sec:methods}, the only difference between Models A and B is the inclusion of radiative cooling in the simulation. Model A therefore represents a three-dimensional numerical test of the original \citetalias{Chevalier85} analytic model with the inclusion of a galactic disk. Similarly, Model B can be directly related to the \cite{Thompson16} outflow model that accounts for radiative energy losses. A comparison between Figures \ref{fig:adiabatic_slices_25} and \ref{fig:cooling_slices_25} shows the dramatic effect that radiative cooling can have on the overall character of a mass-loaded outflow. In these Figures, we plot density\added{, radial velocity,} and temperature slices in the centered $x-z$ plane. Some features are common to both simulations - namely the $10^4\,\mathrm{K}$ disk and the spherical gain region at the center where mass and energy are injected. In Model A, a bi-conical free-wind region is clearly visible in both the density and temperature slices, surrounded by a region of more turbulent, higher temperature gas where the outflow is interacting with the disk gas. In Model B, turbulence at the disk-outflow interface is highly suppressed, \added{leading to a free-wind region with opening angle nearly $\Delta \Omega = 4\pi$. Although energy is injected into the hot wind at the same rate in Models A and B, the energy transferred to the disk gas at the interface is quickly radiated away in Model B. As a result, the interface maintains a much steeper density gradient, which reduces the development of the shear instabilities that drive the turbulence in Model A.} The most obvious feature in the temperature slice for Model B is the clear presence of a cooling radius between 2 and $3\,\mathrm{kpc}$, exactly as predicted by the 1D analytic model.

\begin{figure*}[h!]
\centering
\includegraphics[width=0.33\linewidth]{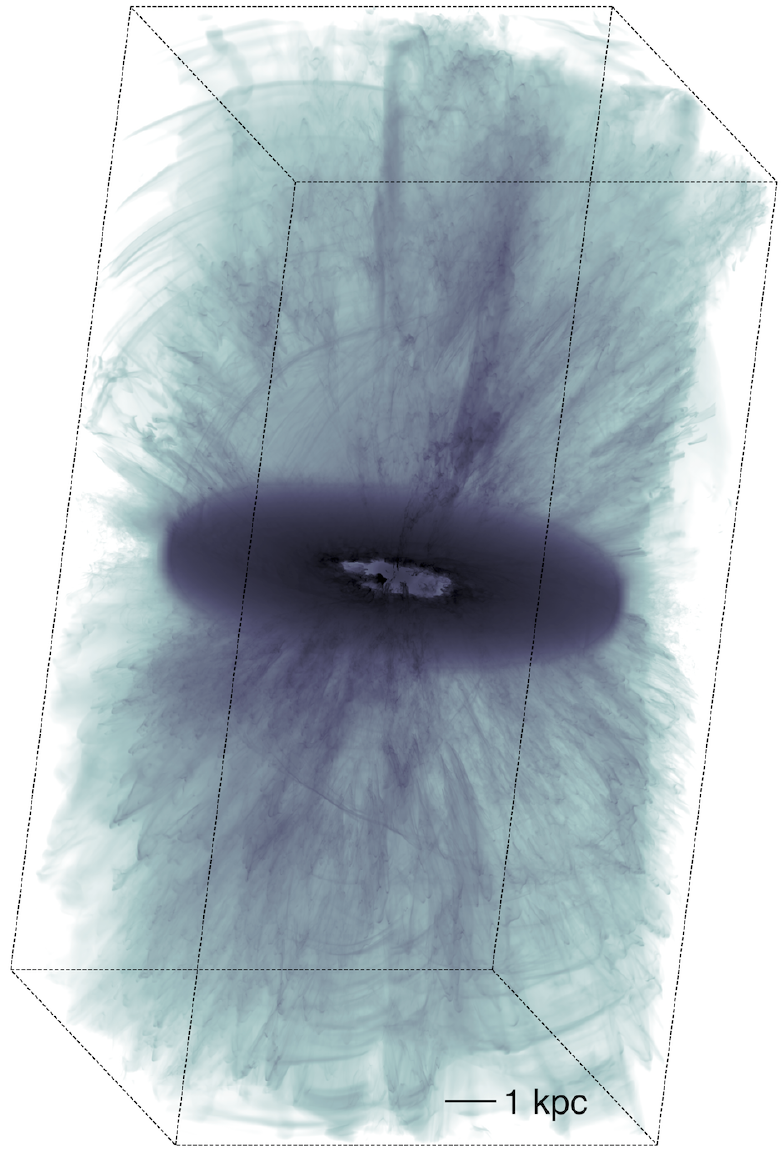}
\includegraphics[width=0.33\linewidth]{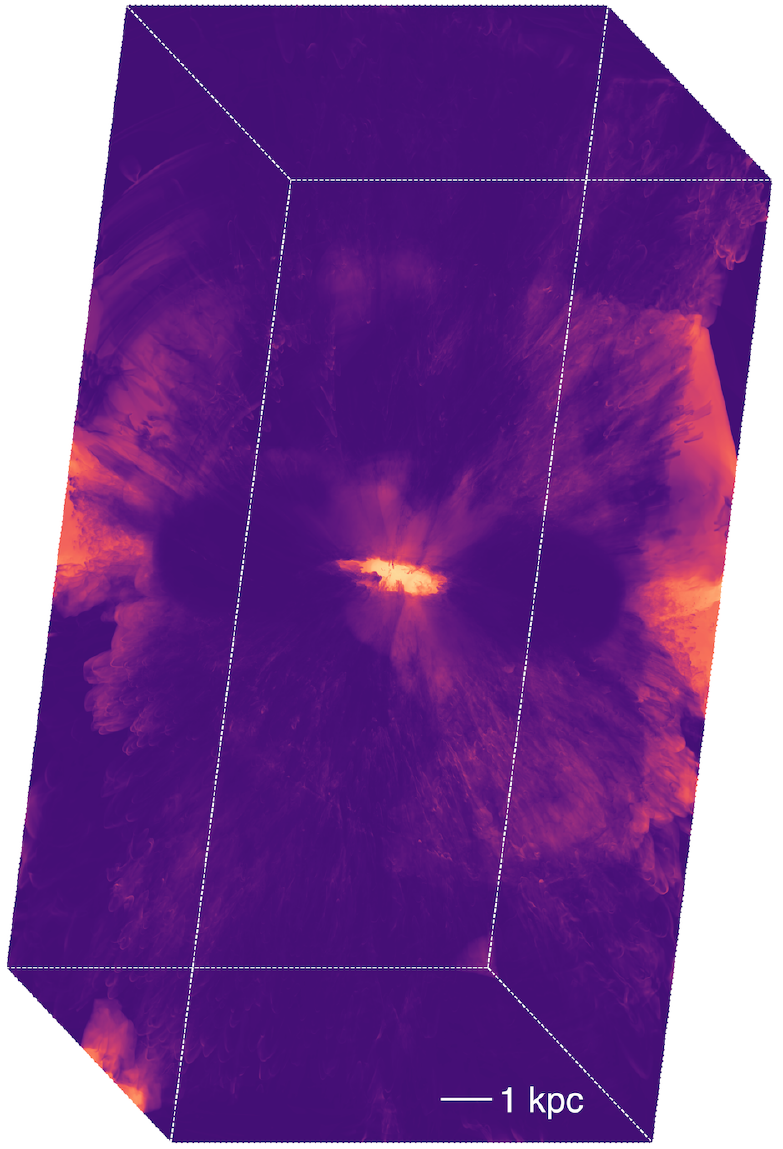}
\caption{Density (left) and density-weighted temperature (right) projections from the clustered feedback simulation (Model C) at $25\,\mathrm{Myr}$, during the high mass-loading outflow state. In contrast with the central feedback simulations, lots of small-scale structure now appears in the outflow. Clouds of high density gas have been lofted above the disk, and spherical shells and  filaments have formed where different outflow solutions interact.}
\label{fig:cluster_projections_25}
\end{figure*}

\begin{figure*}[h!]
\centering
\includegraphics[width=0.3\linewidth]{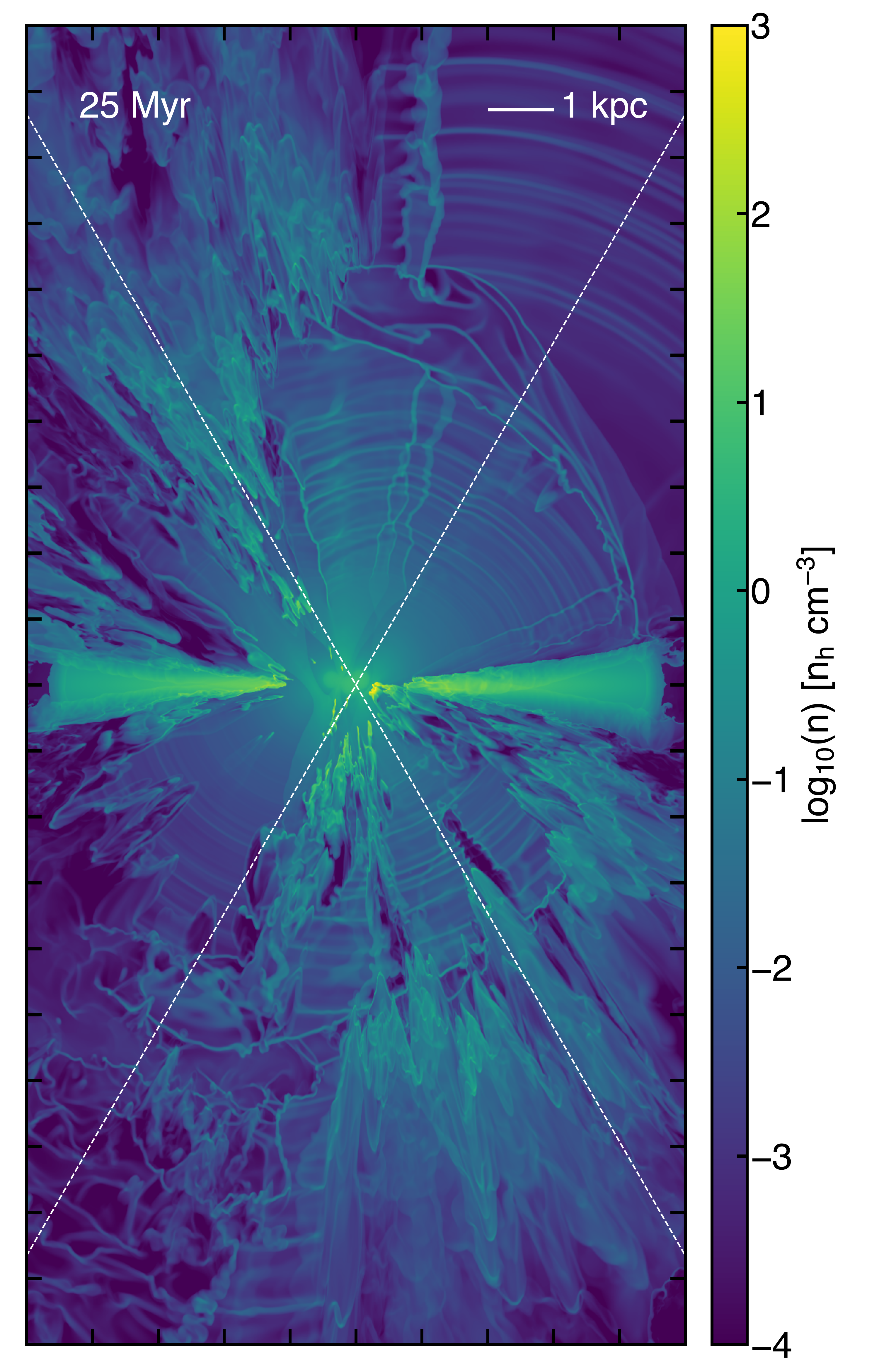}
\includegraphics[width=0.3\linewidth]{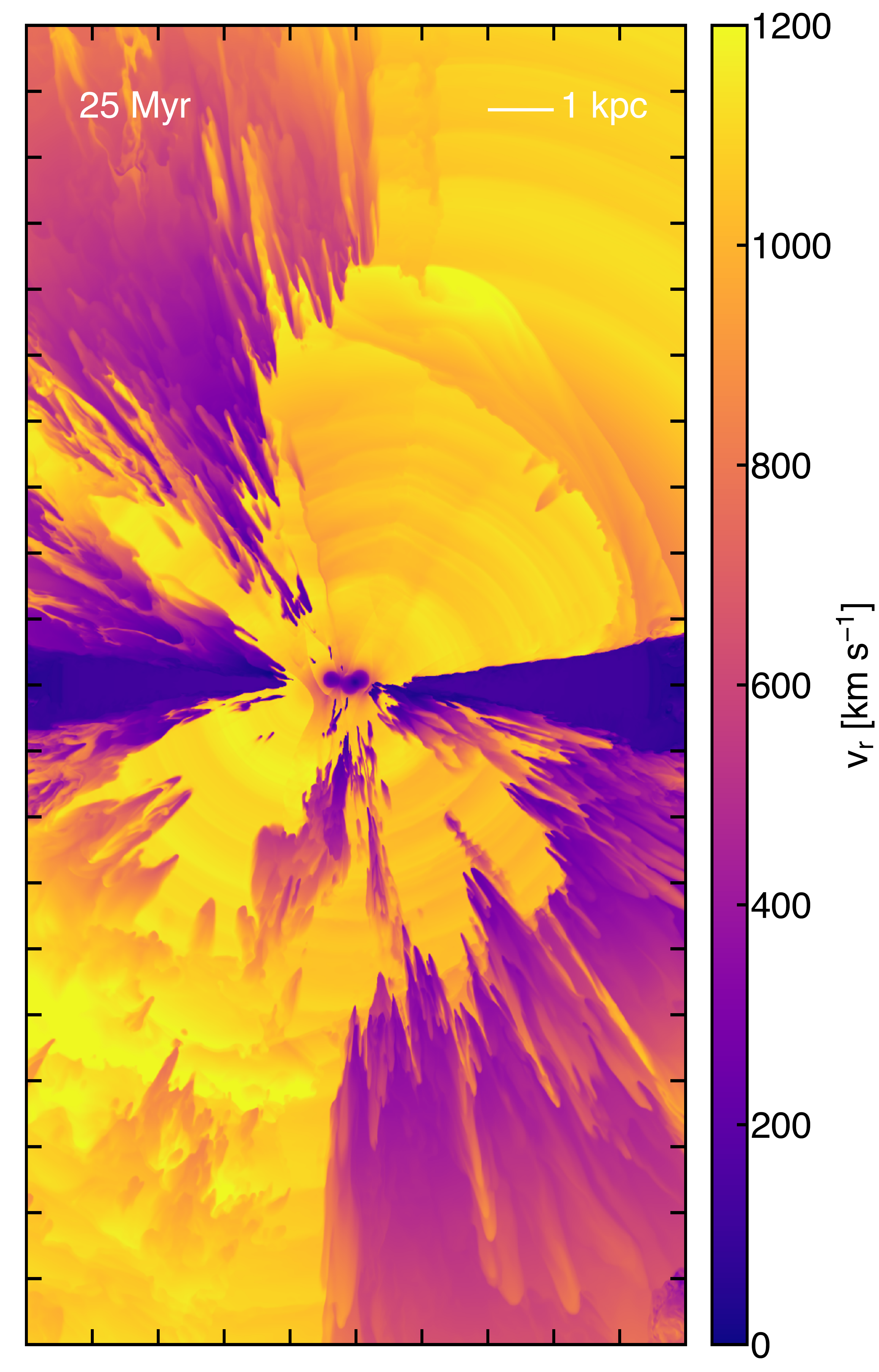}
\includegraphics[width=0.3\linewidth]{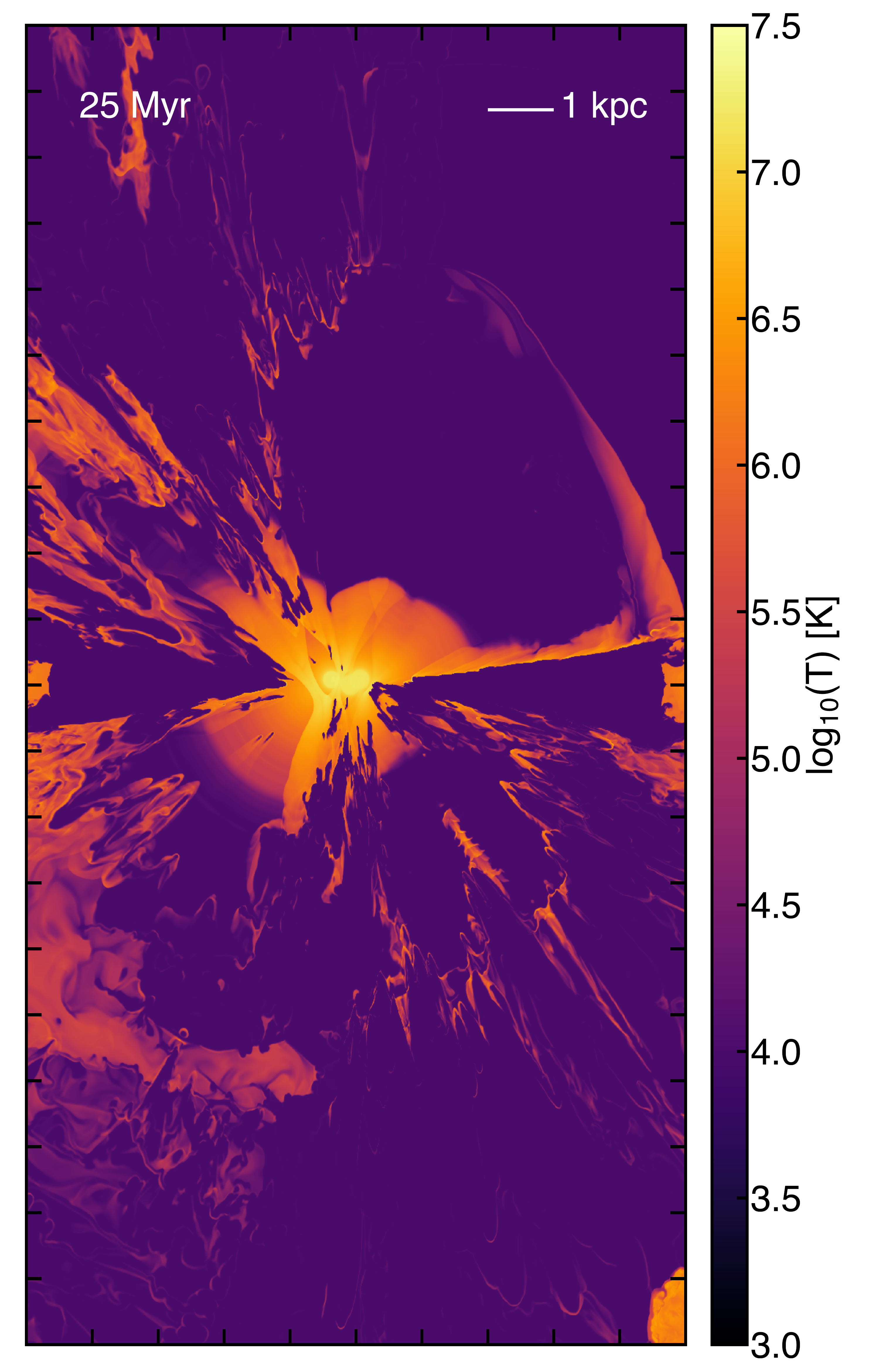}
\caption{Density\added{, radial velocity,} and temperature slices from the clustered feedback simulation shown in Figure \ref{fig:cluster_projections_25}. Large-scale cooling of the hot outflow is still visible in the temperature slice, but there is no longer a single well-defined cooling radius, partly due to the many high-temperature shocks at the interfaces between fast-moving low-density gas and the much more slowly moving high-density gas that has been pushed into the outflow.}
\label{fig:cluster_slices_25}
\end{figure*}

The general character of the solution changes for the clustered feedback Model C. Figure \ref{fig:cluster_projections_25} shows density and temperature projections of the full simulation volume\footnote{Movies showing the full evolution of the clustered feedback simulation from 0 to $75\,\mathrm{Myr}$ can be found online at: https://evaneschneider.org/galactic-wind-simulations/}, while Figure \ref{fig:cluster_slices_25} shows density\added{, radial velocity,} and temperature slices in the $x-z$ plane. The clustered feedback model produces a much more complicated multiphase structure in the outflow as compared to the simulations with axisymmetric central feedback. While large-scale cooling in the outflow is still present, there is no longer a single, well-defined cooling radius, and the density plots display many clouds of higher density, cool gas that have either been lofted out of the disk, or possibly condensed out of the hot wind (see Section~\ref{sec:multiphase}). A number of spherical shells of higher density can also be seen, particularly in the regions with fewer high density clouds. We interpret these shells as interactions between different outflow solutions. Although the energy and mass injection model is constant for each cluster, the rate at which disk gas gets mixed into the hot wind near the disk is highly time-variable, leading to rapidly varying outflow solutions. As a less mass-loaded solution propagates outward, it can overtake a slower-moving, more mass-loaded wind, leading to a series of shells like those seen in the density plot. \added{The clear presence of the shells in the radial velocity slice supports this interpretation.} The relocation of the clusters every $15\,\mathrm{Myr}$ is also correlated with more shell formation, as would be expected when the new outflow regions interact with the old. For example, the large shell at $\sim4 - 5\,\mathrm{kpc}$ seen most clearly in the $25\,\mathrm{Myr}$ temperature slice (Figure~\ref{fig:cluster_slices_25}) was generated when the clusters were relocated at $20\,\mathrm{Myr}$.

\begin{figure*}
\centering
\includegraphics[width=0.3\linewidth]{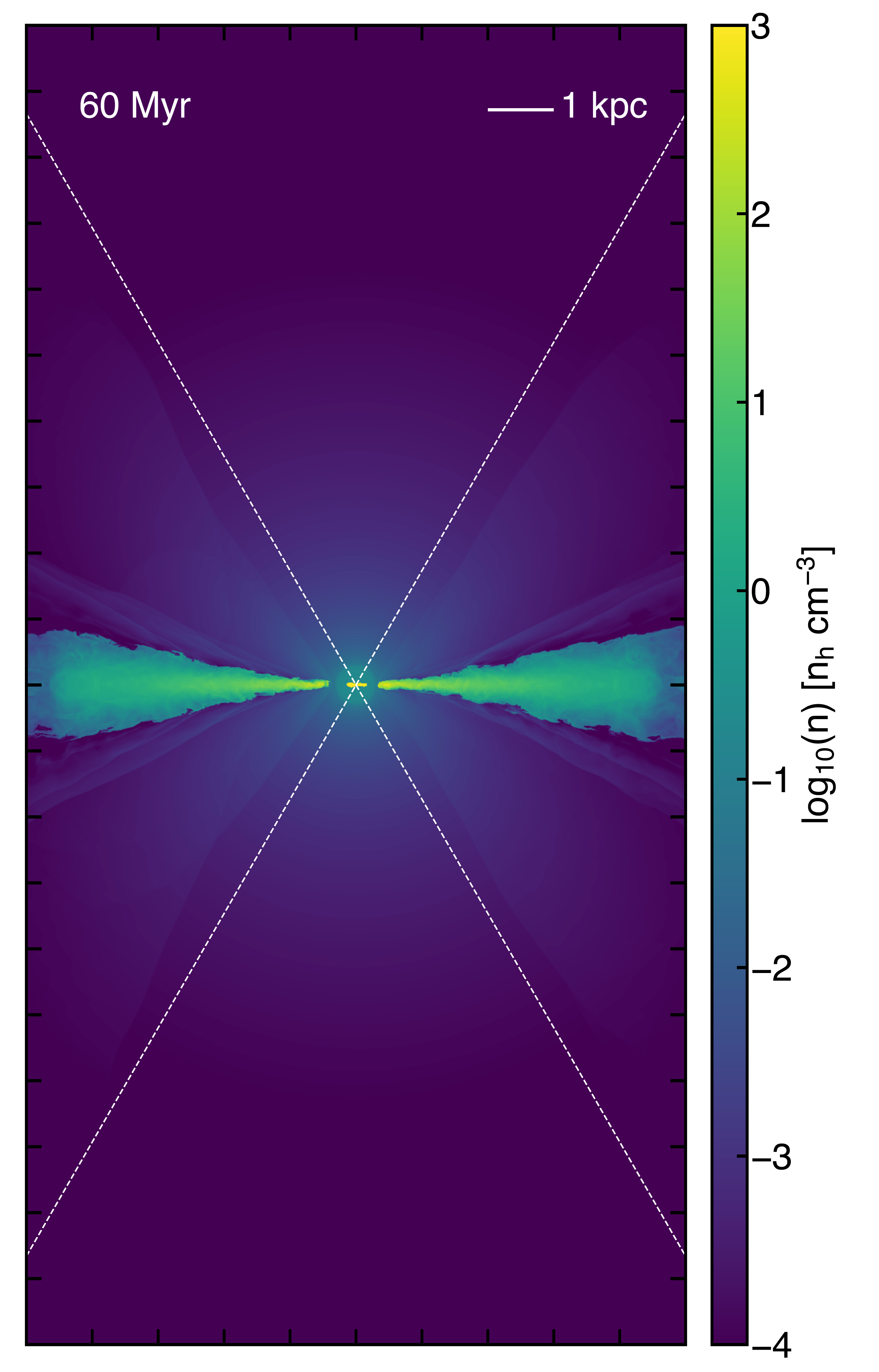}
\includegraphics[width=0.3\linewidth]{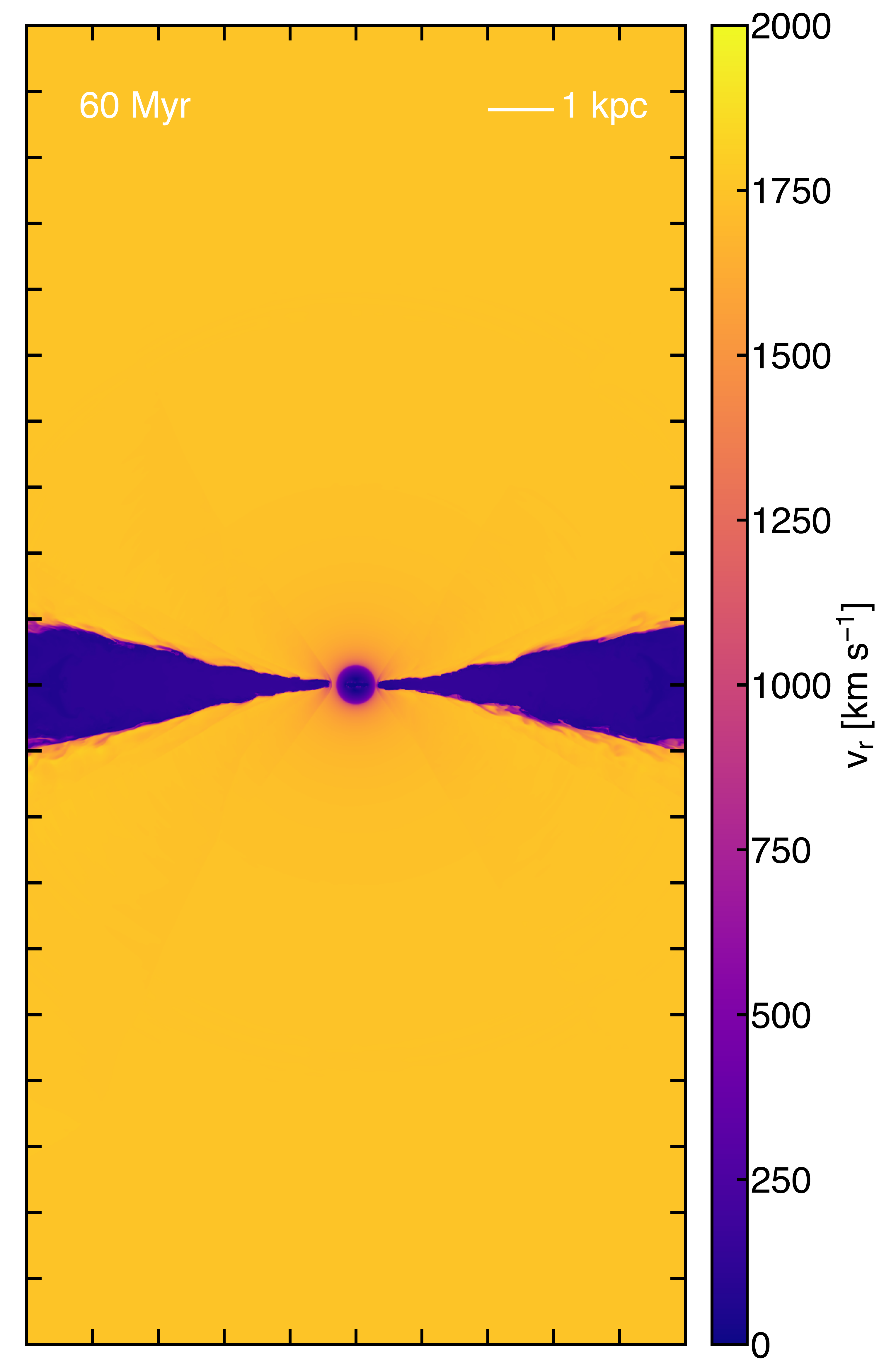}
\includegraphics[width=0.3\linewidth]{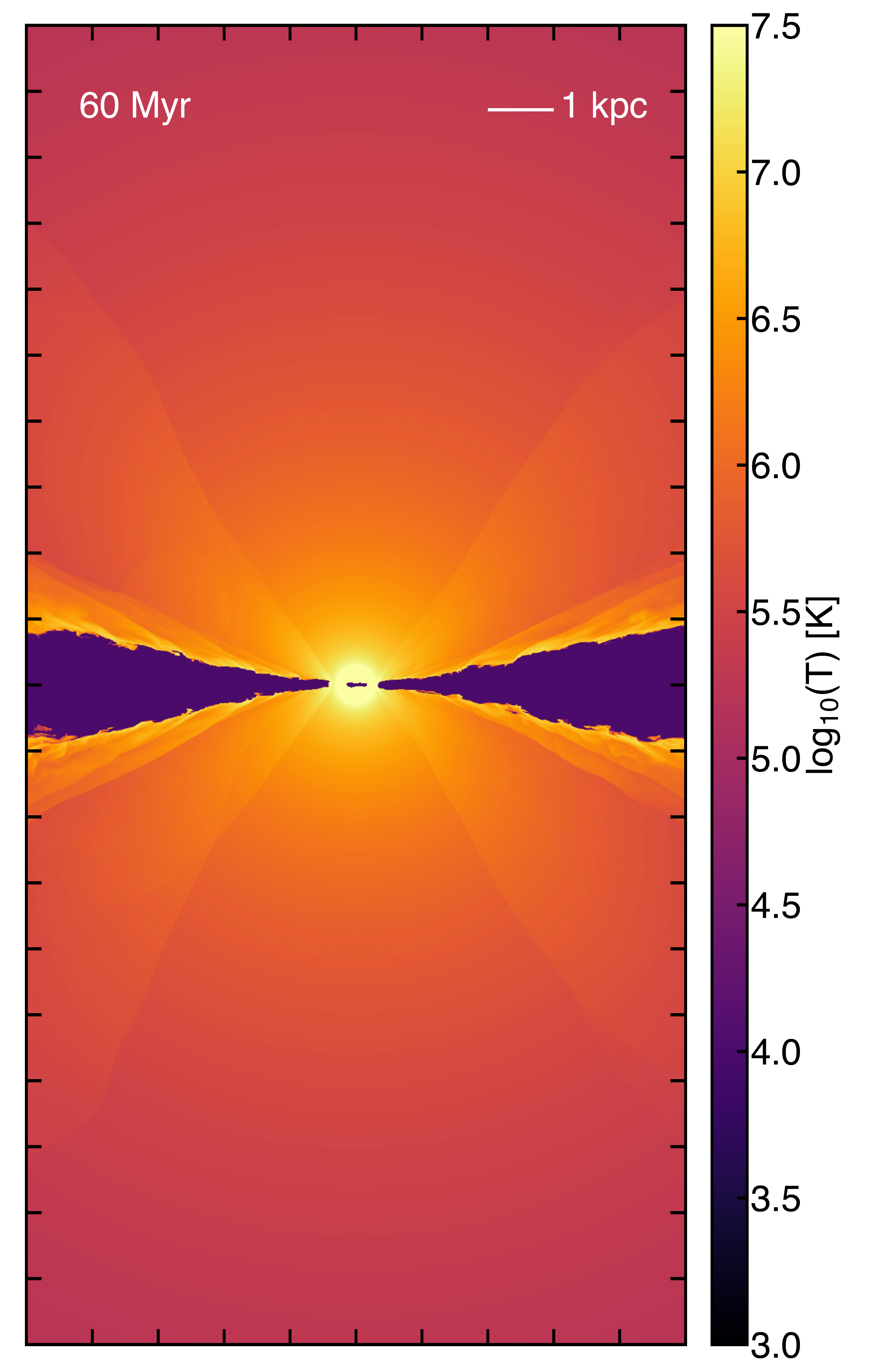}
\caption{Density\added{, radial velocity,} and temperature slices for the central feedback simulation with cooling (Model B) during the low mass-loading state. The hot wind is now too low density for cooling to be efficient at small radii, and the increased speed has also decreased the advection time, so no cooling radius is expected.}
\label{fig:cooling_slices_60}
\end{figure*}

\begin{figure*}
\centering
\includegraphics[width=0.3\linewidth]{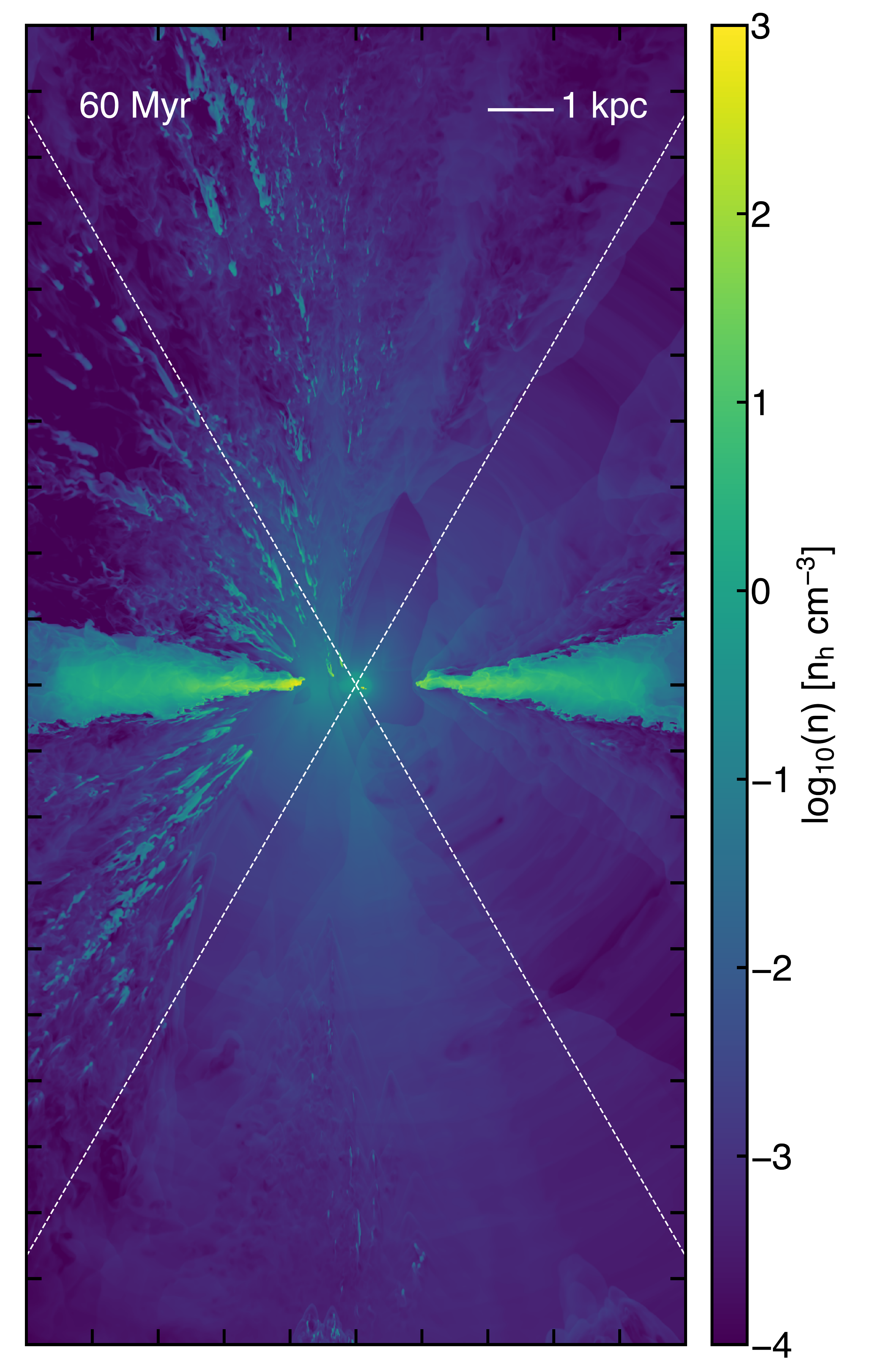}
\includegraphics[width=0.3\linewidth]{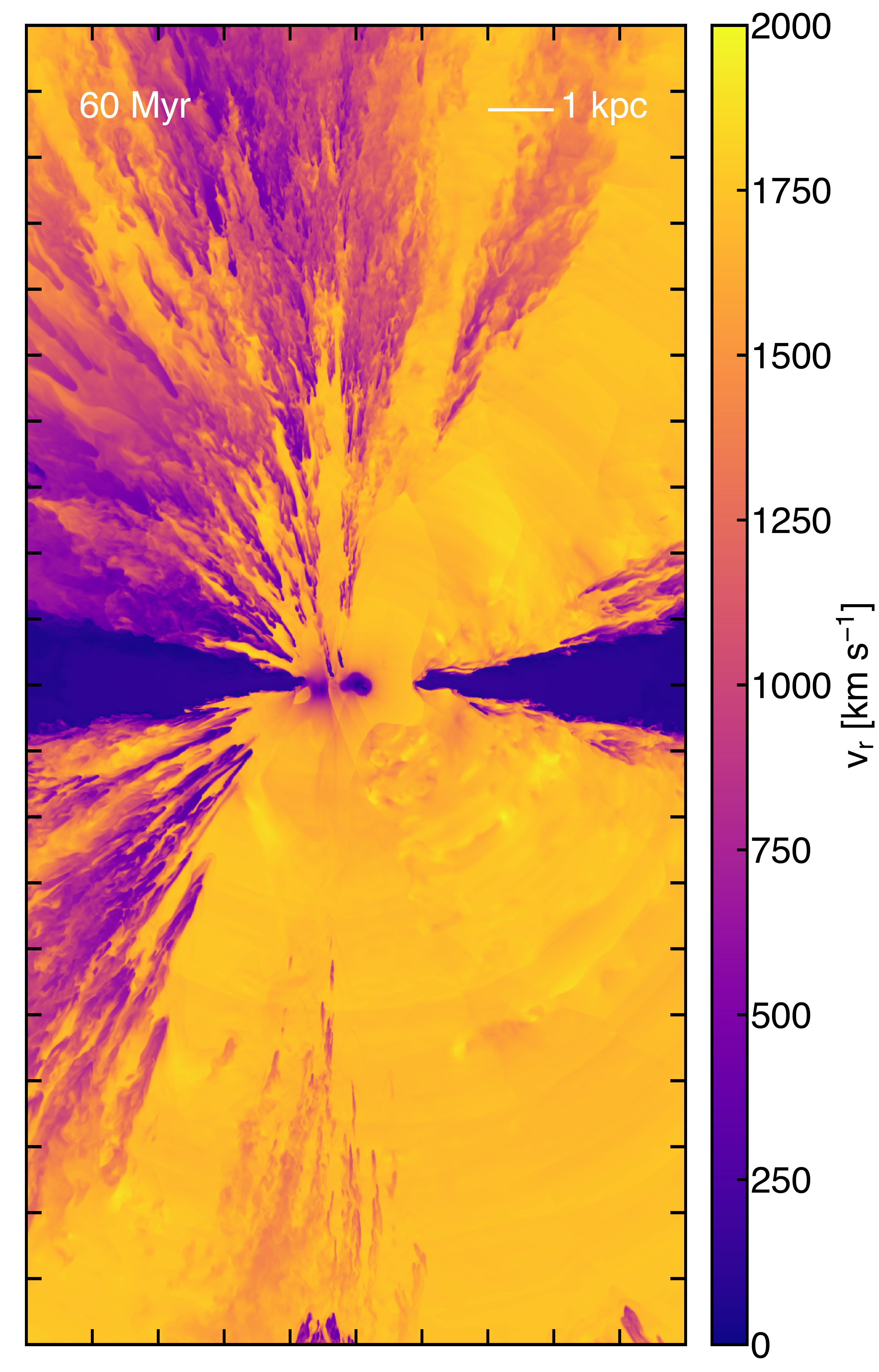}
\includegraphics[width=0.3\linewidth]{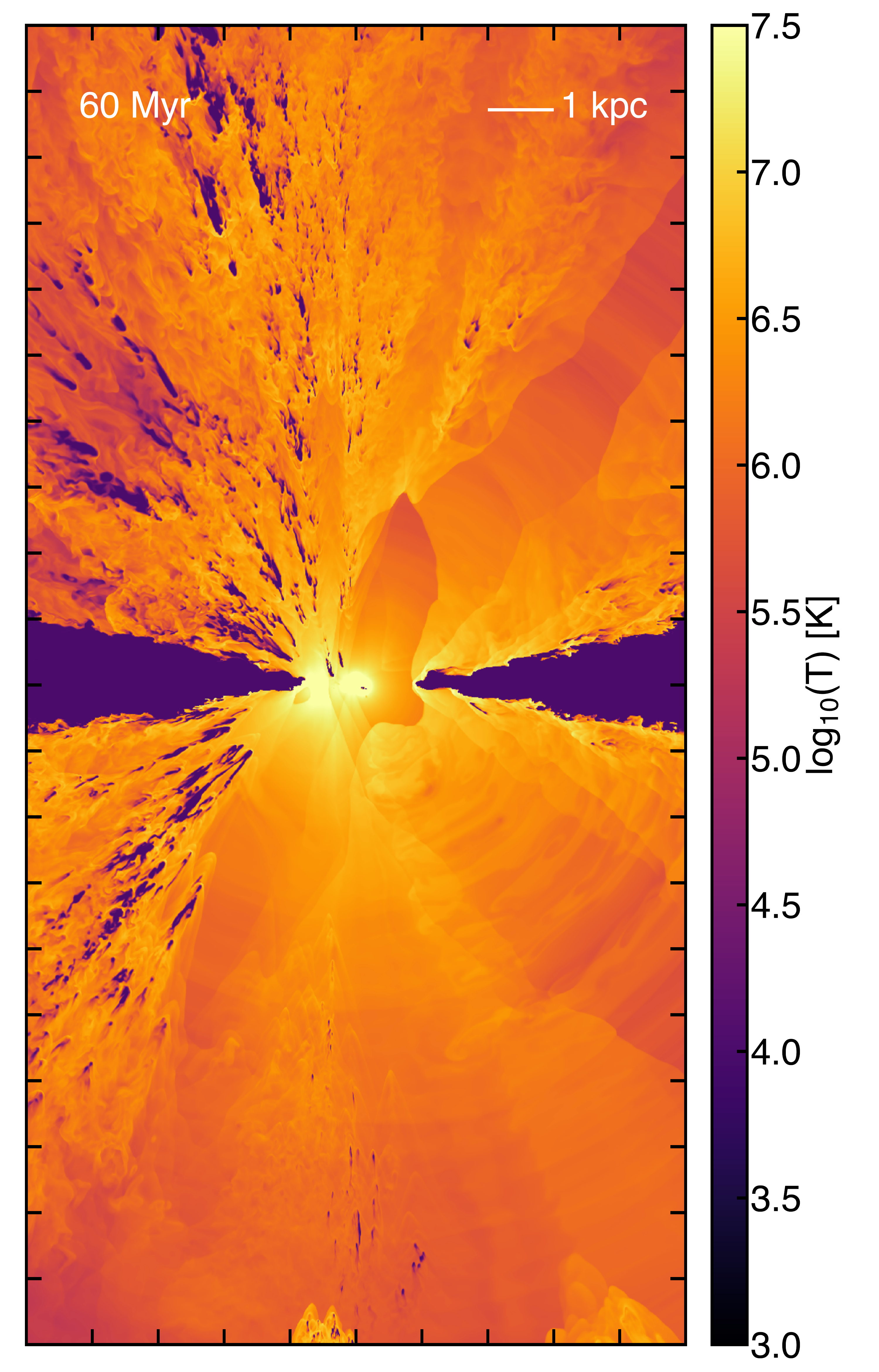}
\caption{Density\added{, radial velocity,} and temperature slices for the clustered feedback simulation (Model C) during the low mass-loading state. As in the central feedback model, the hot wind now remains hot at all radii, but clumps of dense, cool gas that were lofted into the outflow at earlier times remain, and some additional cold material continues to be blown out as rotating disk gas interacts with the clusters.}
\label{fig:cluster_slices_60}
\end{figure*}

As described in Section \ref{sec:methods}, we ramp down the feedback in the simulations between 40 and $45\,\mathrm{Myr}$, allowing us to additionally characterize the outflow during a state with lower mass-loading. We have chosen $60\,\mathrm{Myr}$ as a representative time to display the low mass-loading state, presented for Models B and C in Figures \ref{fig:cooling_slices_60} and \ref{fig:cluster_slices_60}, respectively. We omit additional temperature slices of Model~A because it looks qualitatively similar in both states \citep[though see][for a more detailed discussion of Model A]{Schneider18}. In both cooling models, the background state of the flow has now changed to a hotter, faster wind with no cooling radius, in accordance with the analytic expectation. In fact, Model B now looks very similar to Model A, albeit with much less turbulence at the disk-outflow interface. \added{As a result, the opening angle for the free wind region remains close to $\Delta \Omega = 4\pi$ for Model B even at late times.} The outflow in Model C, however, also contains a series of high temperature bow shocks between the hot wind and the cool, dense material that is entrained in the outflow. This cool material continues to get lofted from the disk into the outflow even at late times in the clustered feedback model, partly because the clusters are moved to a new location within the central $1.5\,\mathrm{kpc}$ of the disk every $15\,\mathrm{Myr}$, and also because the clusters are not rotating with the disk gas, allowing new disk material to enter the cluster sphere of influence and get pushed out as the gas rotates. In addition, the densest clumps of gas that were pushed out during the high mass-loading state are difficult to accelerate, and as a result many of them remain in the simulation volume for millions of years.

\subsection{Radial Properties of the Outflow}\label{sec:radial_structure}

\begin{figure}
\centering
\includegraphics[width=0.95\linewidth]{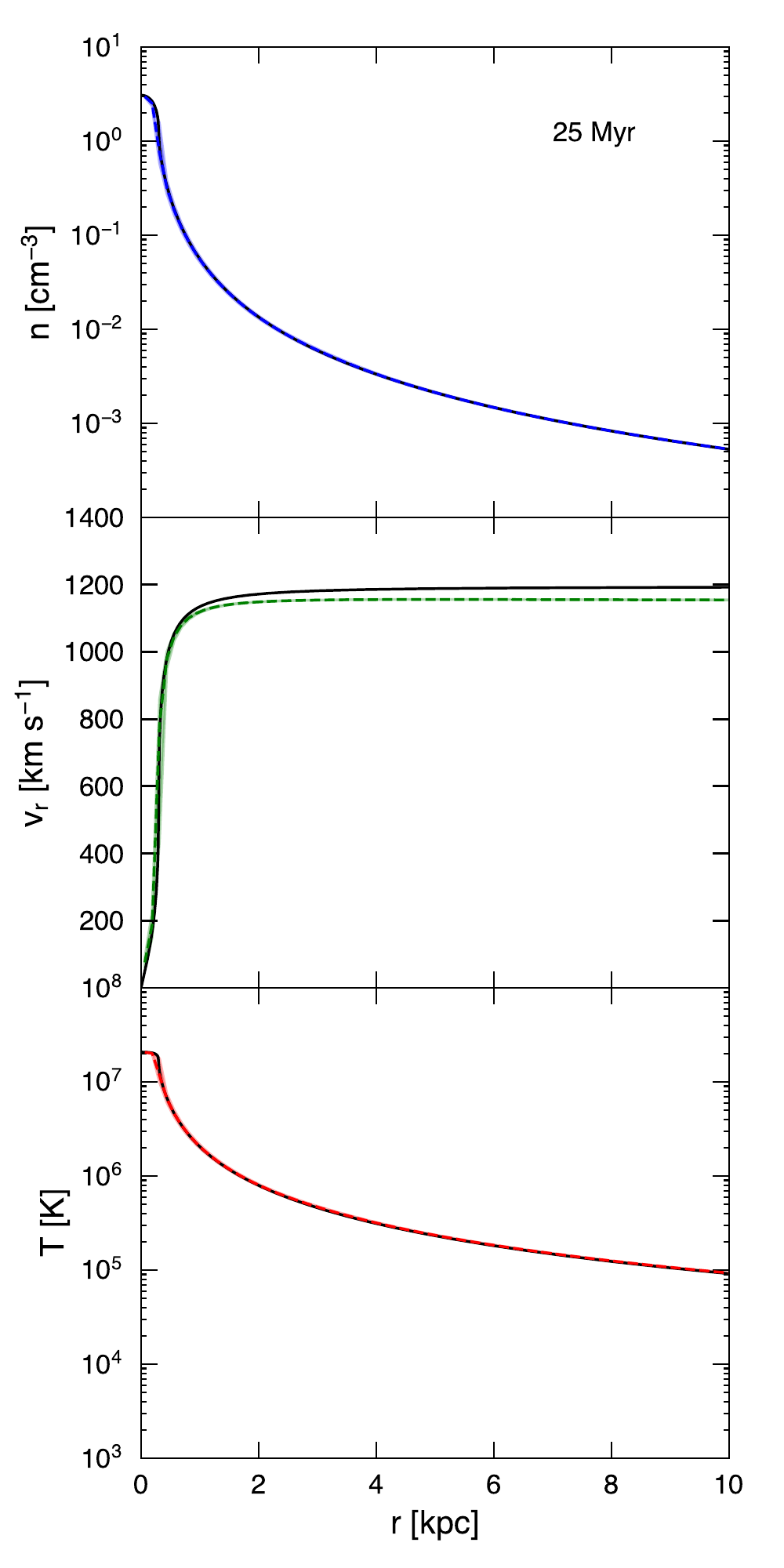}
\caption{\added{Dashed} colored lines show the median density, velocity, and temperature as a function of radius for the central feedback simulation without cooling (Model~A) at $25\,\mathrm{Myr}$. The black solid line shows a 1D numerical calculation of the \citetalias{Chevalier85} solution for each variable\added{, assuming an opening angle of $4\pi\,\mathrm{str}$}. Values for the simulation are calculated in a bi-cone with opening angle $\Delta\Omega = 60^{\circ}$ centered along the $z$-axis\added{, as shown in Figure~\ref{fig:adiabatic_slices_25}}. Upper and lower quartiles within the cone are also plotted; they are identical to the median for this simulation.}
\label{fig:adiabatic_profiles_25}
\end{figure}

In this section we calculate average properties of the outflow as a function of radius and compare them to the solutions from both the \citetalias{Chevalier85} model and the \cite{Thompson16} model.  We start by discussing the adiabatic ``control" simulation, which was designed to mimic the \citetalias{Chevalier85} model while including the effects of a disk and a realistic gravitational potential. Figure \ref{fig:adiabatic_profiles_25} shows radial profiles of Model A at $25\,\mathrm{Myr}$, the same snapshot shown in the density\added{, radial velocity,} and temperature slices from Figure~\ref{fig:adiabatic_slices_25}. We plot three characteristic variables: density, radial velocity (as measured toward the origin), and temperature. In order to focus the results on the outflow, rather than the disk or the interface region, we compute volumetric statistics of these variables for cells within a bi-conical region with an opening angle of $\Delta\Omega = 60^{\circ}$ centered on the $z$-axis, which matches fairly well the opening angle for the free-wind region visible in Figure \ref{fig:adiabatic_slices_25}. \added{The dashed white line in the density panel of Figure~\ref{fig:adiabatic_slices_25} shows the region within which these radial statistics are computed.} Cells are placed into one of 80 equal-size radial bins from $r = 0$ to $r = 10\,\mathrm{kpc}$, and the mean, median, and 25th and 75th percentiles of each variable are calculated for each bin. 

For Model A, all of these statistics are virtually identical \added{within the measured biconical region}, so the lines in Figure~\ref{fig:adiabatic_profiles_25} lie on top of each other, as well as on top of the \citetalias{Chevalier85} solution, plotted in black. Only the radial velocity differs significantly from the analytic model, which is expected as the \citetalias{Chevalier85} model does not include the effect of gravity on the wind. \added{We additionally note that although the slices presented in Figure~\ref{fig:adiabatic_slices_25} show that the free wind region has an opening angle of $\Delta\Omega \sim 60^{\circ}$, the analytic model solution plotted in Figure~\ref{fig:adiabatic_profiles_25} (and subsequent radial plots) uses an opening angle of $4\pi\,\mathrm{str}$, because this angle gives the best match to the undisturbed outflow properties. By definition, our feedback model sets $\Delta\Omega = 4\pi$ at the edge of the injection region, and that appears to be the driving factor in setting the properties of the outflow at larger radii within the free-wind zone.}

\begin{figure}
\centering
\includegraphics[width=0.95\linewidth]{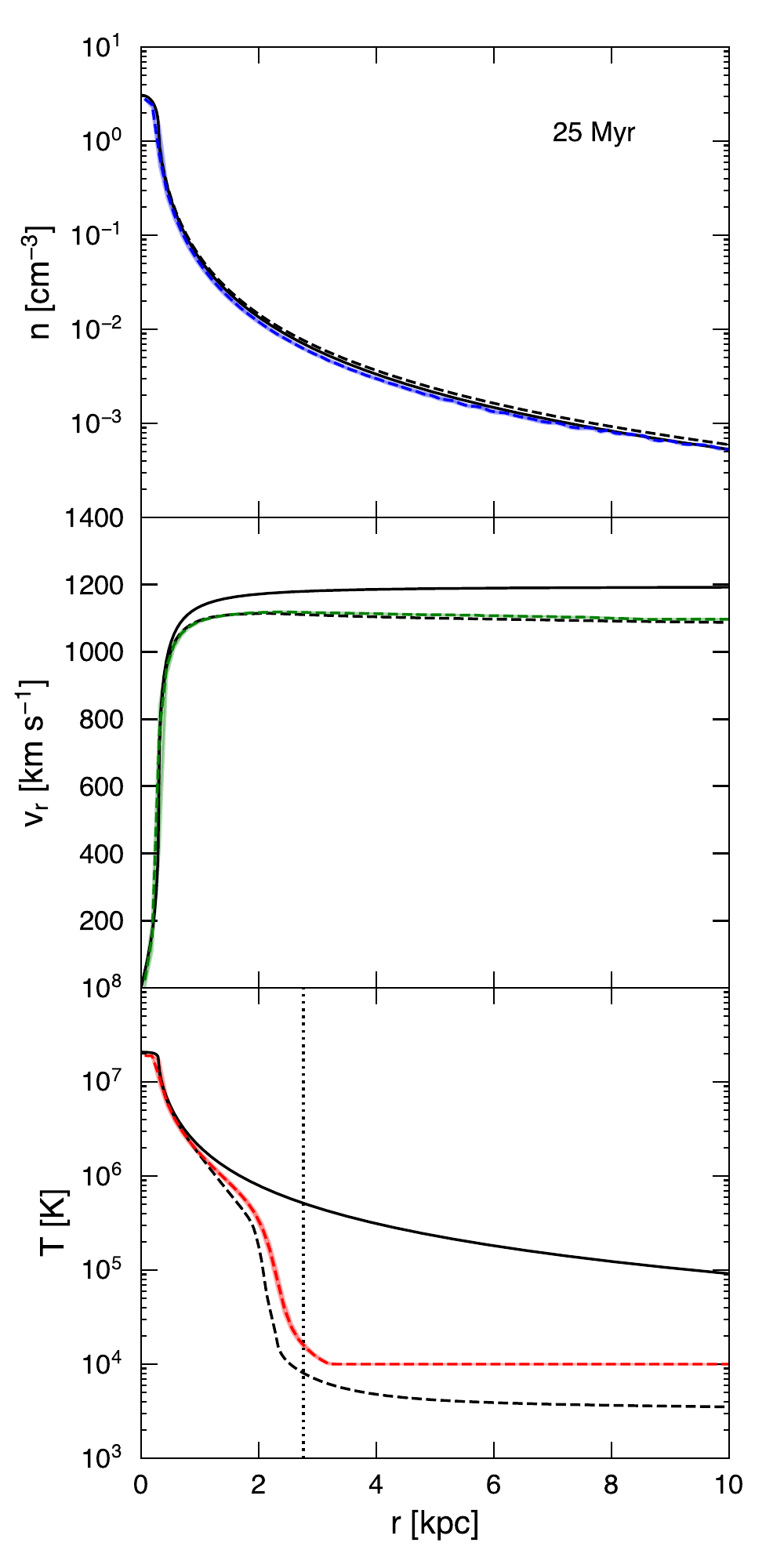}
\caption{Density, velocity, and temperature as a function of radius for the central feedback simulation with cooling (Model B) at $25\,\mathrm{Myr}$. All lines are as in Figure \ref{fig:adiabatic_profiles_25}\added{; upper and lower quartiles again match the median. The additional dashed black line shows a 1D numerical calculation of the \cite{Thompson16} model.} The density and velocity profiles show little signs of change, but the presence of the cooling radius between 2 and $3\,\mathrm{kpc}$ is clear. Gas beyond the cooling radius stays at the simulation temperature floor of $10^4$ K. The analytic estimate of the cooling radius given by Equation~\ref{eqn:r_cool} assuming an opening angle of $4\pi\,\mathrm{str}$ is plotted with the 
\replaced{dashed}{dotted} vertical line.}
\label{fig:cooling_profiles_25}
\end{figure}

The near-perfect match between the \citetalias{Chevalier85} model solution and the Model A simulation provides a convincing proof-of-concept, and gives us a baseline against which to compare the simulations that include the effects of cooling. Figure~\ref{fig:cooling_profiles_25} shows the same three variables plotted against radius for Model B, the central feedback model with cooling. \added{In addition to plotting the \citetalias{Chevalier85} solution (solid black line), we additionally plot the numerically-integrated \cite{Thompson16} outflow solution (dashed black line), which includes both radiative cooling and the effects of the gravitational potential.} The density profile shows almost no change. The velocities are slightly lower, a result of the loss of energy due to radiative cooling before the asymptotic velocity is reached, but the wind still reaches radial velocities over $1000\,\mathrm{km}\,\mathrm{s}^{-1}$. The radially averaged temperature, however, shows a steep decline between 2 and $3\,\mathrm{kpc}$. The exact value of the cooling radius calculated with Equation~\ref{eqn:r_cool} and an outflow opening angle of $\Delta\Omega = 4\pi\,\mathrm{str}$ is plotted with a \replaced{dashed}{dotted} vertical line at $2.8\,\mathrm{kpc}$. Given the various assumptions made in deriving Equation~\ref{eqn:r_cool} (a power-law cooling function, the limit $r >> R$), the agreement between the predicted cooling radius and the radius observed in the simulation is remarkably good.

\begin{figure}
\centering
\includegraphics[width=0.95\linewidth]{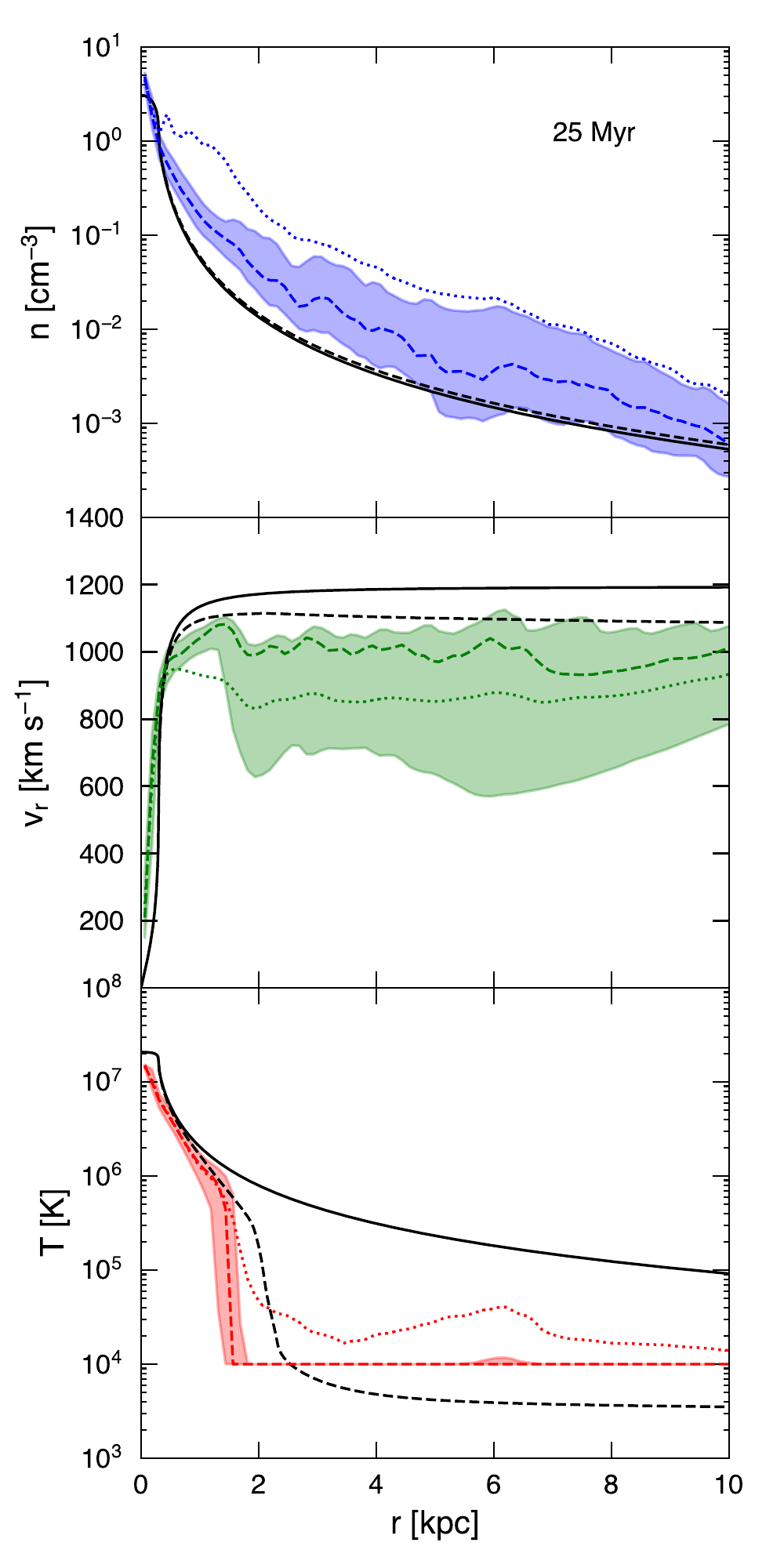}
\caption{Density, velocity, and temperature as a function of radius for the clustered feedback simulation (Model C) at $25\,\mathrm{Myr}$. In addition to the median values (colored dashed lines), we additionally plot the volumetric mean (dotted lines) at all radii. Colored bands show the lower and upper quartile within the $\Delta\Omega = 60^\circ$ cone. Clear enhancements in the density profile reflect the additional entrained disk gas, as do the lowered mean velocities and larger velocity spread. Large-scale cooling of the hot wind is still observed during this high mass-loading state, but the enhanced mean temperature demonstrates that unlike the spherical solution, there is still some hot gas at larger radii in this model.}
\label{fig:cluster_profiles_25}
\end{figure}

The radial profiles for Model C, the clustered feedback simulation, display considerably more complexity. The median density is enhanced by a factor of a few, and even the 25th percentile lies above the analytic solution at almost all radii, reflecting the fact that much of the volume-filling gas has experienced increased mass-loading as a result of mixing between the hot gas injected into the clusters and the cool disk gas. The mean density is above the 75th percentile at nearly all radii, indicating that the outflow also contains clumps of high-density gas that fill a small fraction of the volume but account for a large fraction of the mass. The median radial velocities are decreased relative to the central feedback model, though there is still plenty of cool gas traveling at $v_r > 1000\,\mathrm{km}\,\mathrm{s}^{-1}$. The mean velocity is well below the median at all radii, again reflecting the small volume-filling fraction of very dense gas that has been lofted out of the disk and is traveling at velocities of several hundred $\mathrm{km}\,\mathrm{s}^{-1}$. The radial temperature plot shows that the cooling radius has now moved even closer in compared to the central feedback simulation, while the elevated mean temperature (relative to the median) demonstrates that there is still some hot gas in the outflow at all radii.

\begin{figure}
\centering
\includegraphics[width=0.95\linewidth]{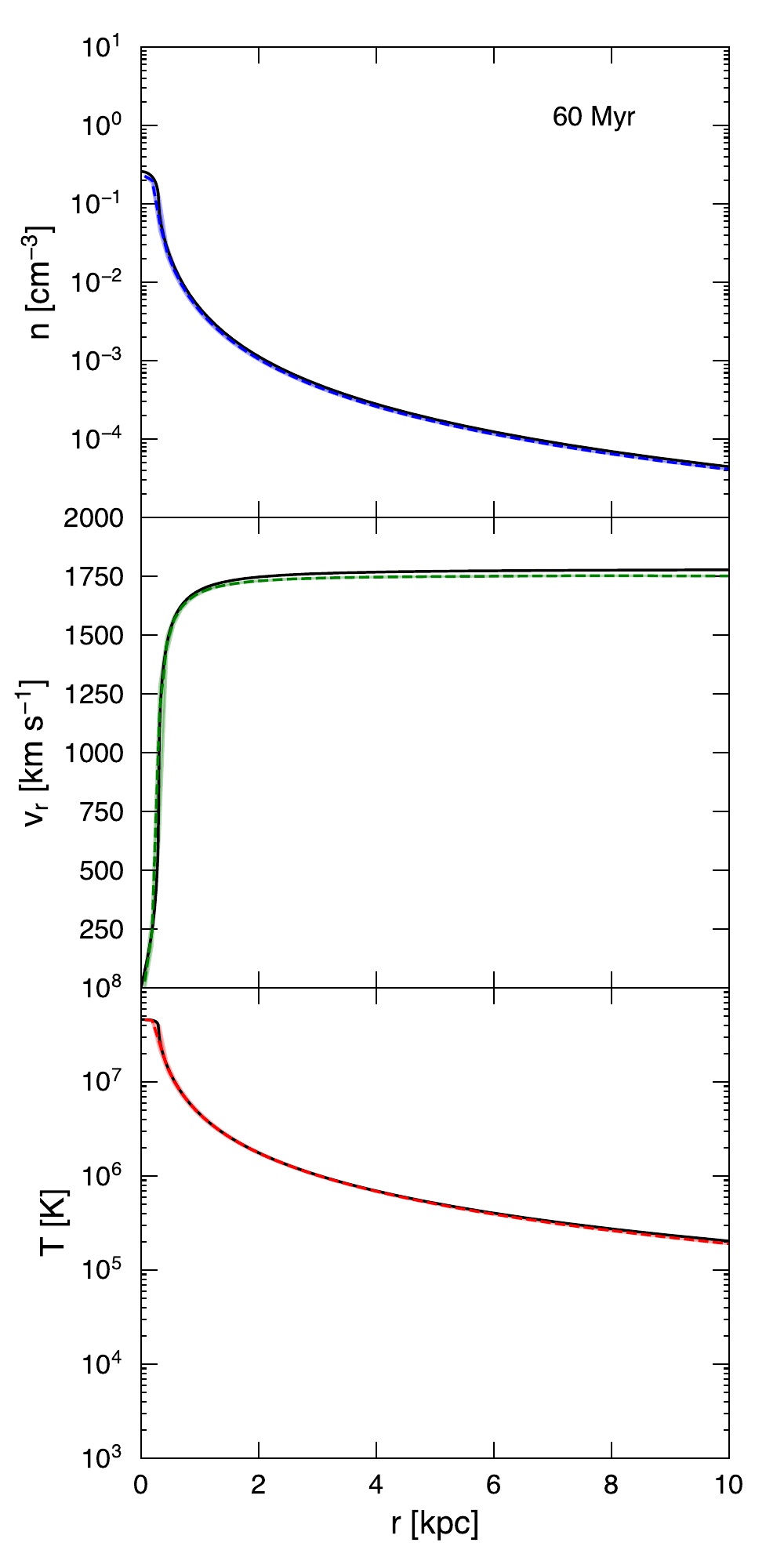}
\caption{Density, velocity, and temperature as a function of radius for the central feedback simulation with cooling (Model B) at $60\,\mathrm{Myr}$, during the low mass-loading outflow state. All lines are as in Figure \ref{fig:adiabatic_profiles_25}. Given the increased cooling time and decreased advection time, the \citetalias{Chevalier85} model now provides an excellent match to the data, even though radiative cooling is included in this simulation.}
\label{fig:cooling_profiles_60}
\end{figure}

\begin{figure}
\centering
\includegraphics[width=0.95\linewidth]{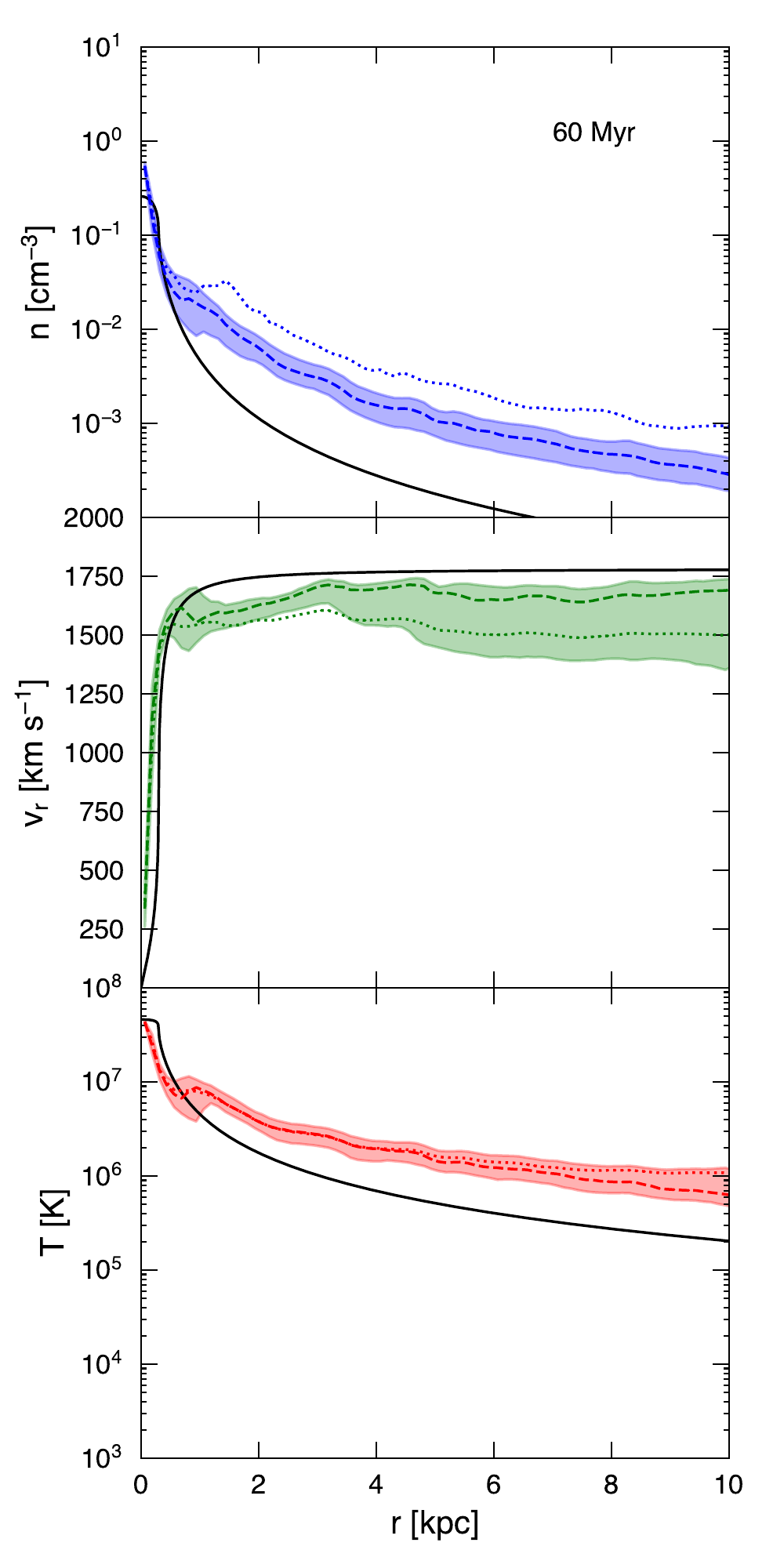}
\caption{Density, velocity, and temperature as a function of radius for the clustered feedback simulation (Model C) at $60\,\mathrm{Myr}$. All lines are as in Figure \ref{fig:cluster_profiles_25}. }
\label{fig:cluster_profiles_60}
\end{figure}

\added{In Figures \ref{fig:cooling_profiles_60} and \ref{fig:cluster_profiles_60} we plot the radial profiles for Models B and C at $60\,\mathrm{Myr}$, during the low mass-loading state. These profiles correspond to the slices shown in Figures \ref{fig:cooling_slices_60} and \ref{fig:cluster_slices_60}. We again omit profiles of Model A at late times due to the similarity between Model A and Model B, but note that the relevant profiles can be found in \citet{Schneider18}. As Figure~\ref{fig:cooling_profiles_60} shows, during the low mass-loading state the \citetalias{Chevalier85} model provides an excellent match to the outflow properties in the central feedback model with cooling. Because the advection time is now much shorter than the cooling time at these radii, radiative losses have little effect on the hot wind, and it escapes the volume with an asymptotic velocity that is almost identical to that predicted by the simple analytic estimate, $v_\mathrm{term} \simeq \sqrt{2\dot{E} / \dot{M}} = 1776\,\mathrm{km}\,\mathrm{s}^{-1}$. The higher ratio of $\alpha$ to $\beta$ in this state means that the velocities and temperatures are higher, while the densities are lower, relative to the high mass-loading state.}

\added{Model C again shows additional complexity relative to the axisymmetric models. As expected based on inspection of Figure~\ref{fig:cluster_slices_60}, the median density of the volume filling gas in the clustered model at late times shows additional mass-loading relative to the analytic solution. Interestingly, while the total amount of mass in the outflow is much lower than at $25\,\mathrm{Myr}$, the mass-loading \textit{relative to what was injected} is considerably higher, with even the 25th percentile over an order of magnitude above the analytic solution at most radii. All three variables - density, velocity, and temperature - show a smaller spread than in the earlier snapshot, as evidenced by the thinner 25th - 75th percentile bands. The elevated mean density continues to indicate the presence of some clumps of higher density gas with a low volume-filling factor. Also interesting is the considerably elevated temperature of the volume-filling gas at all radii at late times, relative to the analytic solution. In \citet{Schneider18} we noted the same elevated temperature profile in Model A at late times, and interpreted it as the result of shocks between the fast-moving hot wind and the slower entrained dense clouds seen in the panels in Figure~\ref{fig:cluster_slices_60}. A rough estimate of the energy produced by slowing the hot wind by $\sim 200\,\mathrm{km}\,\mathrm{s}^{-1}$ (as seen in the velocity profile) gives $\Delta T$ of $\sim 10^6\,\mathrm{K}$, which can explain the increased temperature of the volume-filling hot wind.}

\subsection{Velocity Structure in the Clustered Feedback Simulation}\label{sec:velocity_structure}

\begin{figure*}
\centering
\includegraphics[width=0.9\linewidth]{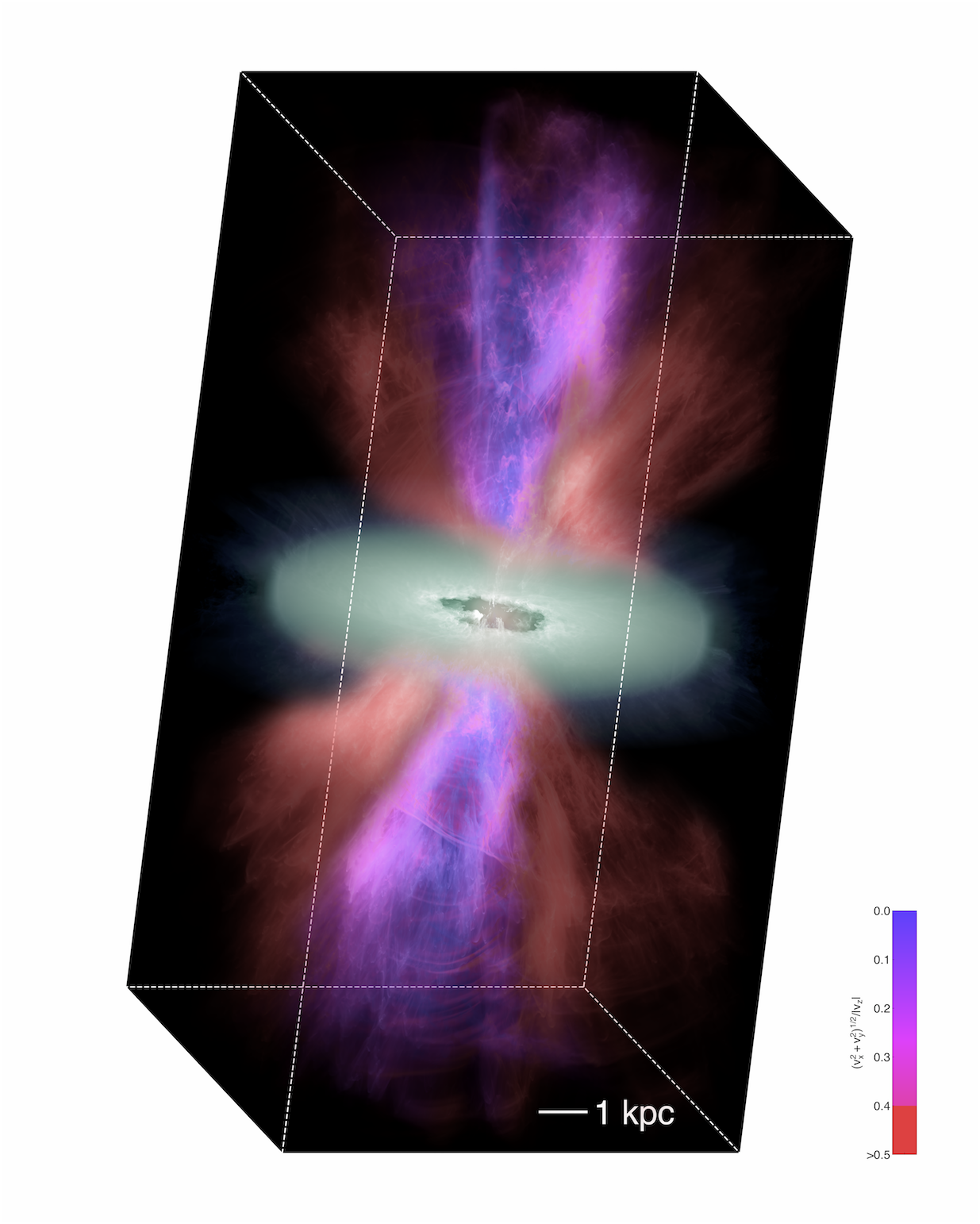}
\caption{Velocity structure of the simulated galaxy outflow from the clustered feedback simulation at $25\,\mathrm{Myr}$. Shown is a logarithmic projected density map (intensity) colorized to highlight the velocity field of the gas. The rotating disk, shown in green, can be kinematically separated from outflowing gas with a large velocity component along the polar $z$-axis (increasing from red to blue). \added{The color bar shows the color mapping for the outflowing gas, which is normalized to the total $z$-velocity. The maximum $z$ velocities occur in gas flowing along the $z$-axis, and are $v \approx 1000\,\mathrm{km}\,\mathrm{s}^{-1}$ (see also Figure \ref{fig:cluster_profiles_25}).}}
\label{fig:cluster_rotation}
\end{figure*}

As evidenced by the plots in the previous sections, a rich amount of density, velocity, and temperature structure exists in the clustered simulation relative to those with spherically symmetric central feedback models. To more clearly visualize this structure, we plot in Figure~\ref{fig:cluster_rotation} a whole-volume realization of the Model C that can be compared to the density and temperature projections in Figure~\ref{fig:cluster_projections_25}, as well as the 1D radial plots in Figure~\ref{fig:cluster_profiles_25}. To create Figure~\ref{fig:cluster_rotation}, we first separate gas into different colormaps depending on the fraction of the velocity that is projected in the $z-direction$. This allows us to cleanly separate out rotating disk gas (plotted in green), versus outflowing gas (plotted in red, pink, and blue). The color separation further shows that the bulk velocity for the majority of the gas is radial, while variations in hue at a given angle from the $z$-axis highlight slight deviations. We additionally set the intensity for a given pixel according to the projected line-of-sight density, which highlights the density enhancements due to the outward-moving shells, as well as the filamentary structures created where outflows from different clusters overlap. The central blue-pink region in Figure~\ref{fig:cluster_rotation} can be directly associated with the $\Delta\Omega = 60^\circ$ bi-conical region used to calculate statistics for the radial profiles. The slower-moving clouds that cause the decreased mean velocity and increased mean density of the profiles in Figure~\ref{fig:cluster_profiles_25} are clearly visible in this central region in Figure~\ref{fig:cluster_rotation}.

\subsection{Multiphase Structure in the Outflow}\label{sec:multiphase}

One goal of this study was to test whether hot, mass-loaded winds could be the source of the multiphase gas seen in many galactic outflows. In this picture, density inhomogeneities in the hot flow, perhaps seeded by the destruction of cool clouds near the base of the wind, are amplified during the cooling phase, resulting in the ``rebirth" of higher density clouds in the outflow at a larger radius. If such a process occurs in winds, the size scale of the high-density clouds when formed was predicted to be small, $<< 1\,\mathrm{kpc}$ \citep[e.g.][]{McCourt18}, which was one of the reasons this study required such high resolution.

\deleted{Unsurprisingly, }We do not see the formation of any such small-scale structures in the simulations using the central feedback model - the constant mass and energy input in the feedback region and relative lack of interaction with the disk gas mean there is no source of density perturbations in the hot wind. \added{Based on the radial profiles shown in Figure \ref{fig:cooling_profiles_25}, there is little or no density enhancement in the wind relative to the injected mass, and the fact that the 25th and 75th percentiles for all variables match the median indicates that there is very little spread in the properties of the volume-filling gas. Rather than small-scale structures seeded by density perturbations cooling out of the hot flow, we find that in the central feedback model, either all the gas cools, or none does.} The clustered model was, in part, an attempt to generate perturbations that could lead to multiphase structure in the outflow. However, the resulting outflow also proved efficient at lofting gas directly from the disk into the wind, making the origin of the resulting multiphase structure difficult to categorize as either accelerated disk gas or cool clouds precipitated from the hot wind.

\begin{figure}[h!]
\centering
\includegraphics[width=0.92\linewidth]{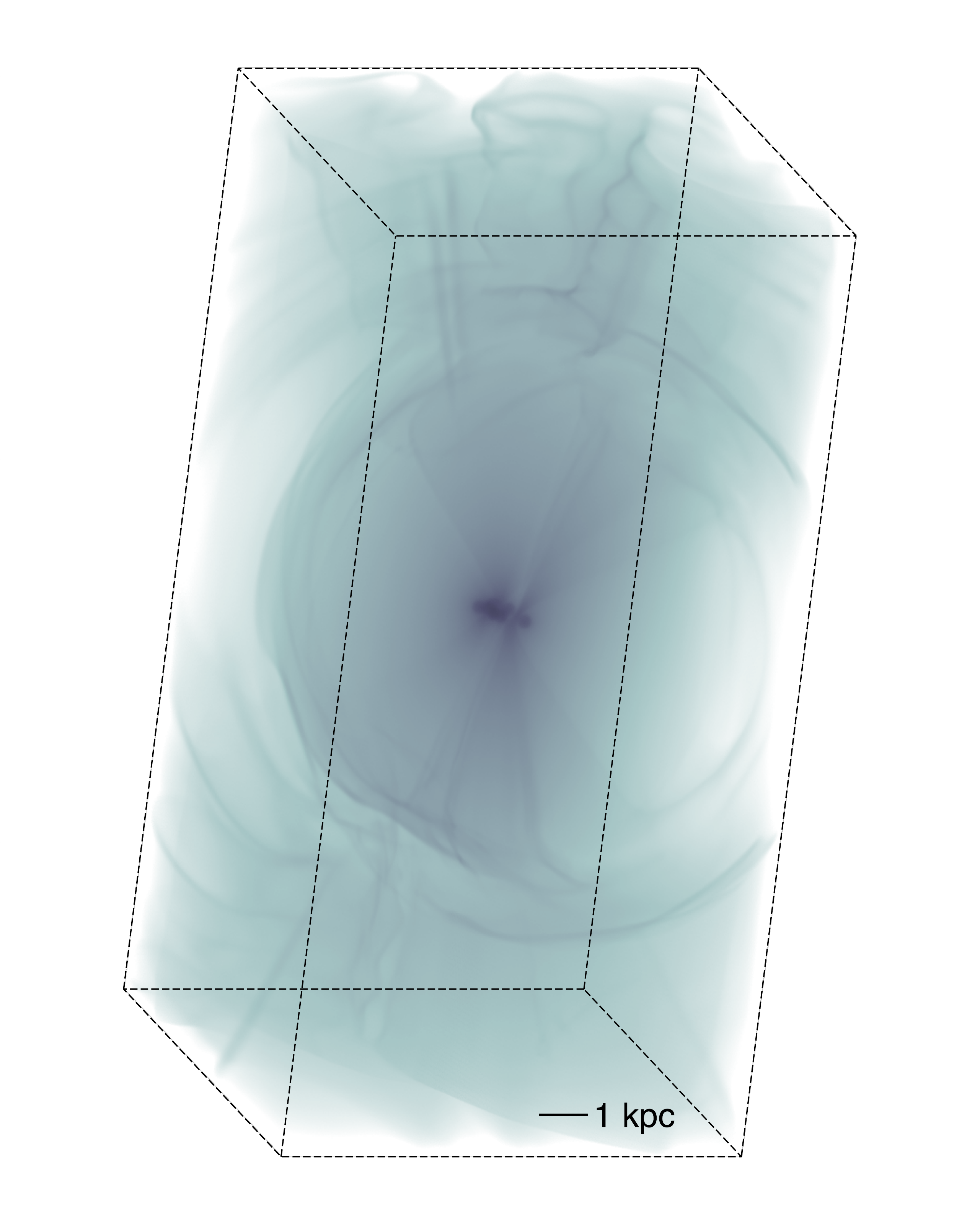}
\includegraphics[width=0.92\linewidth]{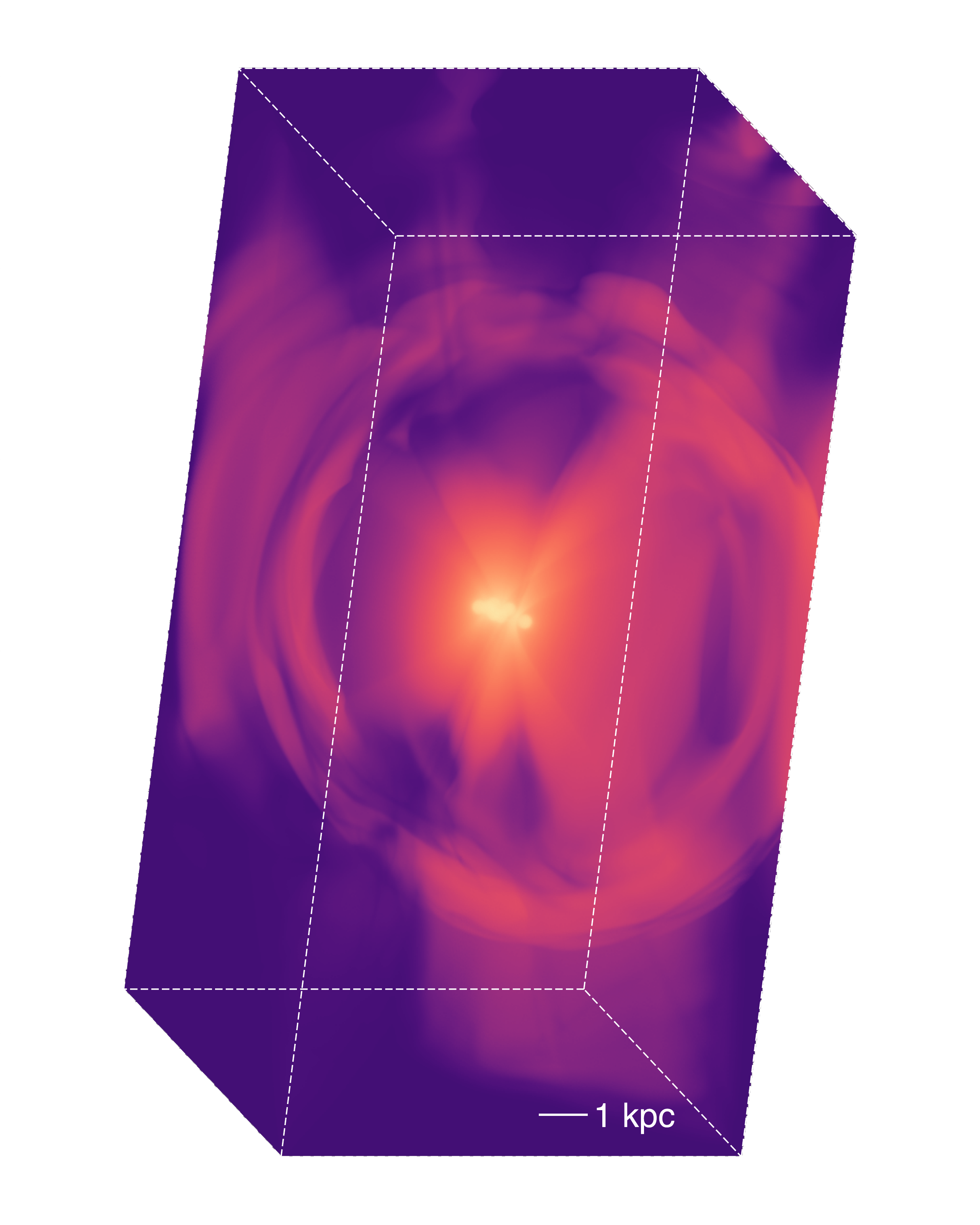}
\caption{Density (top) and density-weighted temperature (bottom) projections from a version of the clustered feedback simulation without a disk. The simulation is shown at $25\,\mathrm{Myr}$, during the high mass-loading outflow state, and can be compared directly with the projections in Figure~\ref{fig:cluster_projections_25}, though note that this simulation has a factor of 4 lower resolution ($\Delta x \approx 20\,\mathrm{pc}$ versus $\Delta x \approx 5\,\mathrm{pc}$.)}
\label{fig:nodisk_projections_25}
\end{figure}

In an attempt to disentangle the origin of the \replaced{higher density}{multiphase} gas in the outflow, we have run a lower-resolution ($\Delta x \approx 20\,\mathrm{pc}$) clustered feedback simulation \textit{without} a disk. The clusters are placed in the same locations as in the high-resolution run, and have the same mass-loading and energy-loading rates. The resulting density and temperature projections at $25\,\mathrm{Myr}$ are shown in Figure~\ref{fig:nodisk_projections_25}. A few features are notable. First, while there are some higher density filaments in the regions where the different cluster outflows overlap, the vast majority of the high density gas from the high-resolution clustered simulation seems to either have been entrained directly from the disk, or condensed out as a result of interactions with the high density clouds in the flow. Second, there are more high-temperature regions in the low-resolution simulation without a disk, and they extend to larger radii. We interpret this as less mass-loading of the hot wind without disk gas to interact with (but we note that resolution also plays a large role in the mass loading of the volume filling phase). Third, the spherical shells visible in both the density and temperature projections appear in both cluster simulations, though there are far more in the higher resolution simulation. This indicates that the shells are caused at least partly by the change in location of the clusters every $15\,\mathrm{Myr}$. Both changes in the location and overlap of the clusters and differences in the mass-loading near the base of the wind change the overall character of the outflow solution, which can then overrun and shock as the wind interacts with outflows from an earlier time. In fact, we do see similar shells in the central feedback model at one particular time - between 40 and $45\,\mathrm{Myr}$, when we are ramping the feedback from the high mass-loading to the low mass-loading state.

To conclude this section, we do not see multiphase outflows being generated directly from the hot phase of the wind in the central feedback simulations. On the other hand, the clustered feedback simulation shows a vast amount of multiphase structure with complex origins, which will be discussed at length in a subsequent paper. These simulations do not rule out the possibility of thermal instability in the hot wind seeding multiphase structure in the outflow, however. The true physical picture may include the generation of outflows with different mass-loading rates from different areas of the disk, and will certainly involve more time variability than our input feedback model. Each of these variables changes the likelihood that some region of the outflow spans the right region of parameter space to generate thermal instability and seed small-scale multiphase structure. A high-resolution, time-variable clustered feedback simulation without a disk could test this theory more rigorously, and is a possibility we may investigate in future work.

\section{Discussion} \label{sec:discussion}

We now turn to a discussion of our simulations in the context of related work on galactic outflows. Although this study is unique in its numerical setup and resolution, many authors have studied galactic outflows via global disk simulations, while recent work by \cite{Scannapieco17} used local-box simulations to address the question of cooling due to thermal instability of mass-loaded winds. We begin by discussing the \cite{Scannapieco17} simulations, as they are the most directly related to this work. Rather than perform an expensive global simulation, Scannapieco performed local box simulations that tracked the outflowing material and included a scale factor designed to mimic the effects of adiabatic expansion. The simulations have high resolution in the co-moving frame ($\Delta x = 0.20\,\mathrm{pc}$), and include detailed modeling of cooling of the outflowing gas, including effects due to atomic/ionic cooling, Compton cooling from scattering of free electrons in the hot wind with photons produced by the galaxy, and dust. By seeding the outflowing material with density perturbations, he is able to track the potential enhancement of perturbations due to cooling, and investigate the possibility of multiphase structure arising in the flow. While the atomic/ionic cooling is found to increase the density fluctuations briefly near the cooling radius, the background medium still cools relatively quickly to the same temperature as the denser clumps, so that on the whole the results are quite similar to those found in our central feedback simulation. At high mass-loading rates comparable to those at early times in our simulations, Scannapieco finds a similar value for the cooling radius, while at lower mass-loading rates (with constant energy loading) the cooling radius moves out or disappears entirely, as predicted \citep{Thompson16}. Thus, neither his simulations nor our central feedback model result in a multiphase outflow at a given radius, but both validate the potential for mass-loaded winds to produce large quantities of cool, fast-moving gas at radii between 1 and $10\,\mathrm{kpc}$.

While not at the scale presented in this work, several previous studies have carried out high-resolution, isolated galaxy simulations with supernova feedback in order to characterize the nature of galactic outflows. In work by \cite{Cooper08} that focused on the wind in M82, the authors injected mass and energy into the starburst region in a manner similar to that used in our central feedback simulations, but their gas disk included a turbulent ISM structure. Due to computational expense, their simulation domain spanned only $1\,\mathrm{kpc}^3$ and ran for only $2\,\mathrm{Myr}$, albeit with a slightly higher physical resolution than that used in our simulations ($\Delta x = 1.9\,\mathrm{pc}$ versus $\Delta x = 4.9\,\mathrm{pc}$). As in our clustered feedback model, they found that cool gas from the disk was effectively launched into the outflow, where it attained velocities ranging from $v = 100 - 800\,\mathrm{km}\,\mathrm{s}^{-1}$ before exiting the simulation volume. While they did not observe cooling of the hot wind, their mass-loading was relatively low: $\beta \sim 0.1$ for a fiducial M82 star formation rate of $10\,\mathrm{M}_{\odot}\,\mathrm{yr}^{-1}$. Additionally, the small volume means that even if the shredding and incorporation of ISM material did increase the mass-loading of the hot phase, the cooling radius would still likely have been outside their simulation volume.

More recent work by \cite{Sarkar15} and \cite{Vijayan18} used a numerical setup similar to ours, with an initially isothermal disk embedded in a hot halo, and injected thermal energy and mass into a centralized or distributed region, respectively. Both works focused on a Milky-Way mass galaxy, and varied the mechanical luminosity of the starburst to focus on the range from $\dot{M}_\mathrm{SFR} = 1.5\,\mathrm{M}_\odot\,\mathrm{yr}^{-1}$ to $\dot{M}_\mathrm{SFR} = 150\,\mathrm{M}_\odot\,\mathrm{yr}^{-1}$. Both studies see some $10^4\,\mathrm{K}$ clouds in the wind, and both trace it to disk gas that has been uplifted by the hot wind, as in our clustered feedback simulation. The general character of both the centralized starburst simulations in \cite{Sarkar15} and the distributed starburst simulations in \cite{Vijayan18} matches our results well. In the centralized case, a conical free-flowing hot wind develops, with a terminal velocity set by the luminosity of the burst ($v_\mathrm{term} \simeq \sqrt{2 \dot{E} / \dot{M}}$) as predicted by the analytic model. Although the authors consider a hot wind mass-loading factor of $\beta = 0.1$, the injection region in their centralized starburst simulations has a radius of only $60\,\mathrm{pc}$, which leads to a high enough star formation rate surface density to see significant fast-moving cool gas in the hot wind on scales ranging from several to tens of kpc. In the distributed case, the star formation rate surface density is not sufficient to drive cooling of the hot phase of the wind, but significantly more of the ISM gas is lofted into the wind, leading to a series of entrained clouds, much like those seen in our clustered feedback simulation.

Many observations have demonstrated the presence of blue-shifted absorption lines in the spectra of starburst galaxies, particularly Mg II. The low ionization potential of the lines indicates that the gas they trace is relatively cool, of order a few $10^4\,\mathrm{K}$, while the maximum velocities measured in the absorption profiles indicate that the gas can achieve high velocity. Measured maximum velocities of Mg II frequently exceed $500\,\mathrm{km}\,\mathrm{s}^{-1}$, and sometimes approach $1000\,\mathrm{km}\,\mathrm{s}^{-1}$ \citep[e.g.][]{Weiner09, Martin09, Rigby17}. While previous global galaxy simulations such as those discussed above have successfully accelerated disk gas to velocities above $500\,\mathrm{km}\,\mathrm{s}^{-1}$, higher resolution studies of cool clouds embedded in hot outflows have indicated that were the resolution in the global simulations high enough to capture the relevant hydrodynamic instabilities, the clouds would likely be shredded and destroyed before achieving the highest velocities observed \citep{Scannapieco15, Bruggen16, Schneider17, Zhang17}, unless magnetic fields can protect them \citep{McCourt15, Banda-Barragan18}. In our clustered feedback simulation, cool gas in the outflow has a dual origin - some has cooled out of the hot phase, while other clouds appear to have been accelerated directly from the disk without undergoing a phase transition. Whether these entrained clouds of disk gas would survive the trip to large radii with higher resolution is unclear. We will investigate the origin of the cool gas in the outflow by including a passive scalar in future simulations. In the meantime, the velocity spread demonstrated in Figure~\ref{fig:cluster_profiles_25} may point toward a satisfactory explanation for the range of observed velocities. While entrained high density clouds near the disk contribute to the low velocity absorption, high velocities can be simultaneously explained by the presence of the fast, cooled wind at larger radii. Geometric effects relative to the line-of-sight could also contribute to the range of velocities observed. The creation of synthetic absorption spectra from the simulations presented here will allow us to better test this model in comparison to the observations, and will be presented in a future work.

\section{Summary and Conclusions} \label{sec:conclusions}

Using the CGOLS suite of extremely high resolution global galaxy simulations, we have highlighted the role of radiative cooling in moderately mass-loaded galactic winds. Our work is complementary to and expands upon the analytic wind models of \cite{Chevalier85}, \cite{Thompson16}, and others. Using a either a central mass and energy injection model or several clustered injection regions to approximate the feedback of supernovae, we vary mass- and energy-loading factors to test for the presence (or absence) of rapid radiative cooling in the hot wind. Our primary conclusions are:
\begin{itemize}
\item{In the adiabatic case with a spherically-symmetric central injection region, the wind model of \cite{Chevalier85} is a good fit to the simulation data within a bi-conical region centered along the minor axis. Outside this region, interactions between the outflow and the disk lead to a turbulent zone with higher density and pressure \citep[see also][]{Schneider18}.}
\item{In the radiative case with central injection, moderate mass-loading of the hot wind within the feedback region ($\beta = 0.6$) leads to rapid cooling at a radius that is accurately predicted by Equation~\ref{eqn:r_cool} \citep[see also][]{Thompson16}. Cooling suppresses turbulence in the region between the disk and outflow, and a lack of density perturbations in the flow means that the outflow is not multiphase in the traditional sense (multiple gas phases at a given radius).}
\item{When the feedback is distributed in a clustered manner throughout the central region of the disk, a truly multiphase outflow is generated. High mass-loading leads to rapid cooling of the hot phase, generating cool gas at velocities of $\sim 1000\,\mathrm{km}\,\mathrm{s}^{-1}$ at radii of $r \sim 2\,\mathrm{kpc}$ and beyond. Significant quantities of cool disk gas are also lofted into the outflow. Interactions between the fast wind and the denser clouds lead to increased mass-loading of the hot phase, which both decreases the cooling radius and decreases the level of mass-loading required in the feedback region in order to trigger rapid cooling in the outflow.}
\end{itemize}

Our results suggest that the cool outflowing gas observed in starburst systems may have a dual origin. Clustered supernova feedback efficiently lofts gas out of the disk and into the hot wind, where it can be accelerated to velocities of several hundred $\mathrm{km}\,\mathrm{s}^{-1}$ before being destroyed. Simultaneously, increased mass-loading of the hot wind as a result of hydrodynamic interactions with this disk gas can lead to rapid cooling of the hot phase, producing cool gas traveling at $500 - 1000\,\mathrm{km}\,\mathrm{s}^{-1}$. This gas will in theory propagate far into the halo, where it could explain observations of cool gas in the CGM, a possibility that we will explore in future work.

\acknowledgments
We acknowledge inspiration from the Simons Symposia on Galactic Superwinds: Beyond Phenomenology.
This research used resources of the Oak Ridge Leadership Computing Facility, which is a DOE Office of Science User Facility supported under Contract DE-AC05-00OR22725, via an award of computing time on the Titan system provided by the INCITE program (AST119, AST125). EES is supported by NASA through Hubble Fellowship grant \#HF-51397.001-A awarded by the Space Telescope Science Institute, which is operated by the Association of Universities for Research in Astronomy, Inc., for NASA, under contract NAS 5-26555. BER acknowledges partial support through NASA contract NNG16PJ25C, and grants 80NSSC18K0563 and HST-GO-14747. TAT is supported in part by NSF \#1516967 and NASA 17-ATP17-0177.

\software{Cholla \citep{Schneider15}; \texttt{matplotlib} \citep{Hunter07}, \texttt{numpy} \citep{VanDerWalt11}, \texttt{hdf5} \citep{hdf5}; Cloudy \citep{Ferland13}}

\bibliography{all}

\begin{thebibliography}{}
\expandafter\ifx\csname natexlab\endcsname\relax\def\natexlab#1{#1}\fi

\bibitem[{{Banda-Barrag{\'a}n} {et~al.}(2018){Banda-Barrag{\'a}n}, {Federrath},
  {Crocker}, \& {Bicknell}}]{Banda-Barragan18}
{Banda-Barrag{\'a}n}, W.~E., {Federrath}, C., {Crocker}, R.~M., \& {Bicknell},
  G.~V. 2018, \mnras, 473, 3454

\bibitem[{{Batten} {et~al.}(1997){Batten}, {Clarke}, {Lambert}, \&
  {Causon}}]{Batten97}
{Batten}, P., {Clarke}, N., {Lambert}, C., \& {Causon}, D.~M. 1997, SIAM
  Journal on Scientific Computing, 18, 1553

\bibitem[{{Bordoloi} {et~al.}(2014){Bordoloi}, {Lilly}, {Hardmeier}, {Contini},
  {Kneib}, {Le Fevre}, {Mainieri}, {Renzini}, {Scodeggio}, {Zamorani},
  {Bardelli}, {Bolzonella}, {Bongiorno}, {Caputi}, {Carollo}, {Cucciati}, {de
  la Torre}, {de Ravel}, {Garilli}, {Iovino}, {Kampczyk}, {Kova{\v c}},
  {Knobel}, {Lamareille}, {Le Borgne}, {Le Brun}, {Maier}, {Mignoli}, {Oesch},
  {Pello}, {Peng}, {Perez Montero}, {Presotto}, {Silverman}, {Tanaka}, {Tasca},
  {Tresse}, {Vergani}, {Zucca}, {Cappi}, {Cimatti}, {Coppa}, {Franzetti},
  {Koekemoer}, {Moresco}, {Nair}, \& {Pozzetti}}]{Bordoloi14}
{Bordoloi}, R., {Lilly}, S.~J., {Hardmeier}, E., {et~al.} 2014, \apj, 794, 130

\bibitem[{{Borthakur} {et~al.}(2016){Borthakur}, {Heckman}, {Tumlinson},
  {Bordoloi}, {Kauffmann}, {Catinella}, {Schiminovich}, {Dav{\'e}}, {Moran}, \&
  {Saintonge}}]{Borthakur16}
{Borthakur}, S., {Heckman}, T., {Tumlinson}, J., {et~al.} 2016, \apj, 833, 259

\bibitem[{{Breitschwerdt} {et~al.}(1991){Breitschwerdt}, {McKenzie}, \&
  {Voelk}}]{Breitschwerdt91}
{Breitschwerdt}, D., {McKenzie}, J.~F., \& {Voelk}, H.~J. 1991, \aap, 245, 79

\bibitem[{{Br{\"u}ggen} \& {Scannapieco}(2016)}]{Bruggen16}
{Br{\"u}ggen}, M., \& {Scannapieco}, E. 2016, \apj, 822, 31

\bibitem[{{Bustard} {et~al.}(2016){Bustard}, {Zweibel}, \&
  {D'Onghia}}]{Bustard16}
{Bustard}, C., {Zweibel}, E.~G., \& {D'Onghia}, E. 2016, \apj, 819, 29

\bibitem[{{Chevalier} \& {Clegg}(1985)}]{Chevalier85}
{Chevalier}, R.~A., \& {Clegg}, A.~W. 1985, \nat, 317, 44

\bibitem[{{Chisholm} {et~al.}(2017){Chisholm}, {Tremonti}, {Leitherer}, \&
  {Chen}}]{Chisholm17}
{Chisholm}, J., {Tremonti}, C.~A., {Leitherer}, C., \& {Chen}, Y. 2017, \mnras,
  469, 4831

\bibitem[{{Chisholm} {et~al.}(2016){Chisholm}, {Tremonti Christy}, {Leitherer},
  \& {Chen}}]{Chisholm16}
{Chisholm}, J., {Tremonti Christy}, A., {Leitherer}, C., \& {Chen}, Y. 2016,
  \mnras, 463, 541

\bibitem[{{Cooper} {et~al.}(2008){Cooper}, {Bicknell}, {Sutherland}, \&
  {Bland-Hawthorn}}]{Cooper08}
{Cooper}, J.~L., {Bicknell}, G.~V., {Sutherland}, R.~S., \& {Bland-Hawthorn},
  J. 2008, \apj, 674, 157

\bibitem[{{Dekel} \& {Silk}(1986)}]{Dekel86}
{Dekel}, A., \& {Silk}, J. 1986, \apj, 303, 39

\bibitem[{{Erb}(2008)}]{Erb08}
{Erb}, D.~K. 2008, \apj, 674, 151

\bibitem[{{Erb} {et~al.}(2012){Erb}, {Quider}, {Henry}, \& {Martin}}]{Erb12}
{Erb}, D.~K., {Quider}, A.~M., {Henry}, A.~L., \& {Martin}, C.~L. 2012, \apj,
  759, 26

\bibitem[{{Ferland} {et~al.}(2013){Ferland}, {Porter}, {van Hoof}, {Williams},
  {Abel}, {Lykins}, {Shaw}, {Henney}, \& {Stancil}}]{Ferland13}
{Ferland}, G.~J., {Porter}, R.~L., {van Hoof}, P.~A.~M., {et~al.} 2013, \rmxaa,
  49, 137

\bibitem[{{Finlator} \& {Dav{\'e}}(2008)}]{Finlator08}
{Finlator}, K., \& {Dav{\'e}}, R. 2008, \mnras, 385, 2181

\bibitem[{{Girichidis} {et~al.}(2016){Girichidis}, {Naab}, {Walch}, {Hanasz},
  {Mac Low}, {Ostriker}, {Gatto}, {Peters}, {W{\"u}nsch}, {Glover}, {Klessen},
  {Clark}, \& {Baczynski}}]{Girichidis16}
{Girichidis}, P., {Naab}, T., {Walch}, S., {et~al.} 2016, \apjl, 816, L19

\bibitem[{{Greco} {et~al.}(2012){Greco}, {Martini}, \& {Thompson}}]{Greco12}
{Greco}, J.~P., {Martini}, P., \& {Thompson}, T.~A. 2012, \apj, 757, 24

\bibitem[{{Heckman} {et~al.}(2017){Heckman}, {Borthakur}, {Wild},
  {Schiminovich}, \& {Bordoloi}}]{Heckman17b}
{Heckman}, T., {Borthakur}, S., {Wild}, V., {Schiminovich}, D., \& {Bordoloi},
  R. 2017, \apj, 846, 151

\bibitem[{{Heckman} {et~al.}(2015){Heckman}, {Alexandroff}, {Borthakur},
  {Overzier}, \& {Leitherer}}]{Heckman15}
{Heckman}, T.~M., {Alexandroff}, R.~M., {Borthakur}, S., {Overzier}, R., \&
  {Leitherer}, C. 2015, \apj, 809, 147

\bibitem[{{Heckman} {et~al.}(2000){Heckman}, {Lehnert}, {Strickland}, \&
  {Armus}}]{Heckman00}
{Heckman}, T.~M., {Lehnert}, M.~D., {Strickland}, D.~K., \& {Armus}, L. 2000,
  \apjs, 129, 493

\bibitem[{{Heckman} \& {Thompson}(2017)}]{Heckman17}
{Heckman}, T.~M., \& {Thompson}, T.~A. 2017, ArXiv e-prints, arXiv:1701.09062

\bibitem[{Hunter(2007)}]{Hunter07}
Hunter, J.~D. 2007, Computing In Science \& Engineering, 9, 90

\bibitem[{{Kim} {et~al.}(2017){Kim}, {Ostriker}, \& {Raileanu}}]{Kim17}
{Kim}, C.-G., {Ostriker}, E.~C., \& {Raileanu}, R. 2017, \apj, 834, 25

\bibitem[{{Kornei} {et~al.}(2012){Kornei}, {Shapley}, {Martin}, {Coil}, {Lotz},
  {Schiminovich}, {Bundy}, \& {Noeske}}]{Kornei12}
{Kornei}, K.~A., {Shapley}, A.~E., {Martin}, C.~L., {et~al.} 2012, \apj, 758,
  135

\bibitem[{{Leitherer} {et~al.}(1999){Leitherer}, {Schaerer}, {Goldader},
  {Delgado}, {Robert}, {Kune}, {de Mello}, {Devost}, \&
  {Heckman}}]{Leitherer99}
{Leitherer}, C., {Schaerer}, D., {Goldader}, J.~D., {et~al.} 1999, \apjs, 123,
  3

\bibitem[{LeVeque(2002)}]{LeVeque02}
LeVeque, R.~J. 2002, Finite-Volume Methods for Hyperbolic Problems (Cambridge
  University Press)

\bibitem[{{Lim} {et~al.}(2013){Lim}, {Hwang}, \& {Lee}}]{Lim13}
{Lim}, S., {Hwang}, N., \& {Lee}, M.~G. 2013, \apj, 766, 20

\bibitem[{{Madau} {et~al.}(2001){Madau}, {Ferrara}, \& {Rees}}]{Madau01}
{Madau}, P., {Ferrara}, A., \& {Rees}, M.~J. 2001, \apj, 555, 92

\bibitem[{{Martin} \& {Bouch{\'e}}(2009)}]{Martin09}
{Martin}, C.~L., \& {Bouch{\'e}}, N. 2009, \apj, 703, 1394

\bibitem[{{Martin} {et~al.}(2012){Martin}, {Shapley}, {Coil}, {Kornei},
  {Bundy}, {Weiner}, {Noeske}, \& {Schiminovich}}]{Martin12}
{Martin}, C.~L., {Shapley}, A.~E., {Coil}, A.~L., {et~al.} 2012, \apj, 760, 127

\bibitem[{{Mashchenko} {et~al.}(2008){Mashchenko}, {Wadsley}, \&
  {Couchman}}]{Mashchenko08}
{Mashchenko}, S., {Wadsley}, J., \& {Couchman}, H.~M.~P. 2008, Science, 319,
  174

\bibitem[{{Mayya} \& {Carrasco}(2009)}]{Mayya09}
{Mayya}, Y.~D., \& {Carrasco}, L. 2009, in Revista Mexicana de Astronomia y
  Astrofisica Conference Series, Vol.~37, Revista Mexicana de Astronomia y
  Astrofisica Conference Series, 44--55

\bibitem[{{Mayya} {et~al.}(2008){Mayya}, {Romano}, {Rodr{\'{\i}}guez-Merino},
  {Luna}, {Carrasco}, \& {Rosa-Gonz{\'a}lez}}]{Mayya08}
{Mayya}, Y.~D., {Romano}, R., {Rodr{\'{\i}}guez-Merino}, L.~H., {et~al.} 2008,
  \apj, 679, 404

\bibitem[{{McCourt} {et~al.}(2018){McCourt}, {Oh}, {O'Leary}, \&
  {Madigan}}]{McCourt18}
{McCourt}, M., {Oh}, S.~P., {O'Leary}, R., \& {Madigan}, A.-M. 2018, \mnras,
  473, 5407

\bibitem[{{McCourt} {et~al.}(2015){McCourt}, {O'Leary}, {Madigan}, \&
  {Quataert}}]{McCourt15}
{McCourt}, M., {O'Leary}, R.~M., {Madigan}, A.-M., \& {Quataert}, E. 2015,
  \mnras, 449, 2

\bibitem[{{Miyamoto} \& {Nagai}(1975)}]{Miyamoto75}
{Miyamoto}, M., \& {Nagai}, R. 1975, \pasj, 27, 533

\bibitem[{{Murray} {et~al.}(2007){Murray}, {Martin}, {Quataert}, \&
  {Thompson}}]{Murray07}
{Murray}, N., {Martin}, C.~L., {Quataert}, E., \& {Thompson}, T.~A. 2007, \apj,
  660, 211

\bibitem[{{Murray} {et~al.}(2005){Murray}, {Quataert}, \&
  {Thompson}}]{Murray05}
{Murray}, N., {Quataert}, E., \& {Thompson}, T.~A. 2005, \apj, 618, 569

\bibitem[{{Navarro} {et~al.}(1996){Navarro}, {Frenk}, \& {White}}]{Navarro96}
{Navarro}, J.~F., {Frenk}, C.~S., \& {White}, S.~D.~M. 1996, \apj, 462, 563

\bibitem[{{Oppenheimer} \& {Dav{\'e}}(2006)}]{Oppenheimer06}
{Oppenheimer}, B.~D., \& {Dav{\'e}}, R. 2006, \mnras, 373, 1265

\bibitem[{{Peeples} \& {Shankar}(2011)}]{Peeples11}
{Peeples}, M.~S., \& {Shankar}, F. 2011, \mnras, 417, 2962

\bibitem[{{Prochaska} {et~al.}(2011){Prochaska}, {Weiner}, {Chen}, {Mulchaey},
  \& {Cooksey}}]{Prochaska11}
{Prochaska}, J.~X., {Weiner}, B., {Chen}, H.-W., {Mulchaey}, J., \& {Cooksey},
  K. 2011, \apj, 740, 91

\bibitem[{{Rigby} {et~al.}(2017){Rigby}, {Bayliss}, {Chisholm}, {Bordoloi},
  {Sharon}, {Gladders}, {Johnson}, {Paterno-Mahler}, {Wuyts}, {Dahle}, \&
  {Acharyya}}]{Rigby17}
{Rigby}, J.~R., {Bayliss}, M.~B., {Chisholm}, J., {et~al.} 2017, ArXiv
  e-prints, arXiv:1710.07499

\bibitem[{{Rubin} {et~al.}(2014){Rubin}, {Prochaska}, {Koo}, {Phillips},
  {Martin}, \& {Winstrom}}]{Rubin14}
{Rubin}, K.~H.~R., {Prochaska}, J.~X., {Koo}, D.~C., {et~al.} 2014, \apj, 794,
  156

\bibitem[{{Sarkar} {et~al.}(2015){Sarkar}, {Nath}, {Sharma}, \&
  {Shchekinov}}]{Sarkar15}
{Sarkar}, K.~C., {Nath}, B.~B., {Sharma}, P., \& {Shchekinov}, Y. 2015, \mnras,
  448, 328

\bibitem[{{Scannapieco}(2017)}]{Scannapieco17}
{Scannapieco}, E. 2017, \apj, 837, 28

\bibitem[{{Scannapieco} \& {Br{\"u}ggen}(2015)}]{Scannapieco15}
{Scannapieco}, E., \& {Br{\"u}ggen}, M. 2015, \apj, 805, 158

\bibitem[{{Scannapieco} {et~al.}(2002){Scannapieco}, {Ferrara}, \&
  {Madau}}]{Scannapieco02}
{Scannapieco}, E., {Ferrara}, A., \& {Madau}, P. 2002, \apj, 574, 590

\bibitem[{{Schneider} \& {Robertson}(2015)}]{Schneider15}
{Schneider}, E.~E., \& {Robertson}, B.~E. 2015, \apjs, 217, 24

\bibitem[{{Schneider} \& {Robertson}(2017)}]{Schneider17}
---. 2017, \apj, 834, 144

\bibitem[{{Schneider} \& {Robertson}(2018)}]{Schneider18}
---. 2018, \apj, 860, 135

\bibitem[{{Silich} {et~al.}(2004){Silich}, {Tenorio-Tagle}, \&
  {Rodr{\'{\i}}guez-Gonz{\'a}lez}}]{Silich04}
{Silich}, S., {Tenorio-Tagle}, G., \& {Rodr{\'{\i}}guez-Gonz{\'a}lez}, A. 2004,
  \apj, 610, 226

\bibitem[{{Socrates} {et~al.}(2008){Socrates}, {Davis}, \&
  {Ramirez-Ruiz}}]{Socrates08}
{Socrates}, A., {Davis}, S.~W., \& {Ramirez-Ruiz}, E. 2008, \apj, 687, 202

\bibitem[{{Steidel} {et~al.}(2010){Steidel}, {Erb}, {Shapley}, {Pettini},
  {Reddy}, {Bogosavljevi{\'c}}, {Rudie}, \& {Rakic}}]{Steidel10}
{Steidel}, C.~C., {Erb}, D.~K., {Shapley}, A.~E., {et~al.} 2010, \apj, 717, 289

\bibitem[{{Stone} \& {Gardiner}(2009)}]{Stone09}
{Stone}, J.~M., \& {Gardiner}, T. 2009, \na, 14, 139

\bibitem[{{Strickland} \& {Heckman}(2009)}]{Strickland09}
{Strickland}, D.~K., \& {Heckman}, T.~M. 2009, \apj, 697, 2030

\bibitem[{{The HDF Group}(1997-NNNN)}]{hdf5}
{The HDF Group}. 1997-NNNN, {Hierarchical Data Format, version 5}, , ,
  http://www.hdfgroup.org/HDF5/

\bibitem[{{Thompson} {et~al.}(2015){Thompson}, {Fabian}, {Quataert}, \&
  {Murray}}]{Thompson15}
{Thompson}, T.~A., {Fabian}, A.~C., {Quataert}, E., \& {Murray}, N. 2015,
  \mnras, 449, 147

\bibitem[{{Thompson} {et~al.}(2016){Thompson}, {Quataert}, {Zhang}, \&
  {Weinberg}}]{Thompson16}
{Thompson}, T.~A., {Quataert}, E., {Zhang}, D., \& {Weinberg}, D.~H. 2016,
  \mnras, 455, 1830

\bibitem[{{Toro} {et~al.}(1994){Toro}, {Spruce}, \& {Speares}}]{Toro94}
{Toro}, E.~F., {Spruce}, M., \& {Speares}, W. 1994, Shock Waves, 4, 25

\bibitem[{{Tremonti} {et~al.}(2004){Tremonti}, {Heckman}, {Kauffmann},
  {Brinchmann}, {Charlot}, {White}, {Seibert}, {Peng}, {Schlegel}, {Uomoto},
  {Fukugita}, \& {Brinkmann}}]{Tremonti04}
{Tremonti}, C.~A., {Heckman}, T.~M., {Kauffmann}, G., {et~al.} 2004, \apj, 613,
  898

\bibitem[{{Tumlinson} {et~al.}(2017){Tumlinson}, {Peeples}, \&
  {Werk}}]{Tumlinson17}
{Tumlinson}, J., {Peeples}, M.~S., \& {Werk}, J.~K. 2017, \araa, 55, 389

\bibitem[{{Tumlinson} {et~al.}(2013){Tumlinson}, {Thom}, {Werk}, {Prochaska},
  {Tripp}, {Katz}, {Dav{\'e}}, {Oppenheimer}, {Meiring}, {Ford}, {O'Meara},
  {Peeples}, {Sembach}, \& {Weinberg}}]{Tumlinson13}
{Tumlinson}, J., {Thom}, C., {Werk}, J.~K., {et~al.} 2013, \apj, 777, 59

\bibitem[{{Van Der Walt} {et~al.}(2011){Van Der Walt}, {Colbert}, \&
  {Varoquaux}}]{VanDerWalt11}
{Van Der Walt}, S., {Colbert}, S.~C., \& {Varoquaux}, G. 2011, Computing in
  Science and Engineering, 13, 22

\bibitem[{{Veilleux} {et~al.}(2005){Veilleux}, {Cecil}, \&
  {Bland-Hawthorn}}]{Veilleux05}
{Veilleux}, S., {Cecil}, G., \& {Bland-Hawthorn}, J. 2005, \araa, 43, 769

\bibitem[{{Vijayan} {et~al.}(2018){Vijayan}, {Sarkar}, {Nath}, {Sharma}, \&
  {Shchekinov}}]{Vijayan18}
{Vijayan}, A., {Sarkar}, K.~C., {Nath}, B.~B., {Sharma}, P., \& {Shchekinov},
  Y. 2018, \mnras, 475, 5513

\bibitem[{{Wang}(1995)}]{Wang95}
{Wang}, B. 1995, \apj, 444, 590

\bibitem[{{Weiner} {et~al.}(2009){Weiner}, {Coil}, {Prochaska}, {Newman},
  {Cooper}, {Bundy}, {Conselice}, {Dutton}, {Faber}, {Koo}, {Lotz}, {Rieke}, \&
  {Rubin}}]{Weiner09}
{Weiner}, B.~J., {Coil}, A.~L., {Prochaska}, J.~X., {et~al.} 2009, \apj, 692,
  187

\bibitem[{{Werk} {et~al.}(2014){Werk}, {Prochaska}, {Tumlinson}, {Peeples},
  {Tripp}, {Fox}, {Lehner}, {Thom}, {O'Meara}, {Ford}, {Bordoloi}, {Katz},
  {Tejos}, {Oppenheimer}, {Dav{\'e}}, \& {Weinberg}}]{Werk14}
{Werk}, J.~K., {Prochaska}, J.~X., {Tumlinson}, J., {et~al.} 2014, \apj, 792, 8

\bibitem[{{Werk} {et~al.}(2016){Werk}, {Prochaska}, {Cantalupo}, {Fox},
  {Oppenheimer}, {Tumlinson}, {Tripp}, {Lehner}, \& {McQuinn}}]{Werk16}
{Werk}, J.~K., {Prochaska}, J.~X., {Cantalupo}, S., {et~al.} 2016, \apj, 833,
  54

\bibitem[{{Zhang} {et~al.}(2018){Zhang}, {Davis}, {Jiang}, \&
  {Stone}}]{Zhang18}
{Zhang}, D., {Davis}, S.~W., {Jiang}, Y.-F., \& {Stone}, J.~M. 2018, \apj, 854,
  110

\bibitem[{{Zhang} {et~al.}(2017){Zhang}, {Thompson}, {Quataert}, \&
  {Murray}}]{Zhang17}
{Zhang}, D., {Thompson}, T.~A., {Quataert}, E., \& {Murray}, N. 2017, \mnras,
  468, 4801

\end{thebibliography}

\listofchanges

\end{document}